\documentclass[prd,10pt,aps,twocolumn,floatfix,amssymb,amsmath,nofootinbib]{revtex4-1}
\usepackage{slashed}

\usepackage{epsfig}
\newcommand{\beqn}{\begin{eqnarray}}
\newcommand{\eeqn}{\end{eqnarray}}
\newcommand{\eq}[1]{(\ref{#1})}
\newcommand{\RE}{{\mathrm{Re}}}
\newcommand{\IM}{{\mathrm{Im}}}
\newcommand{\cL}{{\cal L}}

\newcommand{\cE}{{\cal E}}

\newcommand{\Z}{{\mathbb Z}}
\newcommand{\N}{{\mathbb N}}

\begin{document}

\title{Fractal energy carpets in non--Hermitian Hofstadter quantum mechanics}

\author{M.~N.~Chernodub}\email{maxim.chernodub@lmpt.univ-tours.fr}
\affiliation{CNRS, Laboratoire de Math\'ematiques et Physique Th\'eorique UMR 7350, Universit\'e de Tours, 37200 France}
\affiliation{Department of Physics and Astronomy, University of Gent, Krijgslaan 281, S9, B-9000 Gent, Belgium}
\affiliation{Soft Matter Physics Laboratory, Far Eastern Federal University, Sukhanova str., 8, Vladivostok, 690950, Russia}

\author{St\'ephane Ouvry}\email{stephane.ouvry@u-psud.fr}
\affiliation{CNRS, Laboratoire de Physique Th\'eorique et Mod\`eles Statistiques UMR  8626, Universit\'e Paris-Sud, 91405 Orsay, France}

\begin{abstract}
We study the non-Hermitian Hofstadter dynamics of a quantum particle with  biased motion on a square lattice in the background of a magnetic field. We show that in  quasi-momentum space the energy spectrum  is an overlap of infinitely many inequivalent fractals. The energy levels in each fractal are space-filling curves with  Hausdorff dimension 2. The band structure of the spectrum is similar to a fractal spider net in  contrast to the Hofstadter  butterfly for  unbiased motion.

\end{abstract}


\date{April 9, 2015}

\maketitle

\section{Introduction}

It is well known that the energy spectrum of an electrically charged particle moving in the background of an external magnetic magnetic field on a two-dimensional infinite square lattice has a beautiful fractal structure known as the Hofstadter butterfly~\cite{ref:Hofstadter}. The energy bands plotted against  the  magnetic flux reveal a complex pattern which resembles a butterfly, hence the name ``butterfly''. The fractality of the Hofstadter's butterfly is revealed in the fact that small regions of the energy spectrum contain a distorted copy of a larger region, thus exhibiting the self-similarity property at all scales. This property is a characteristic feature of a fractal~\cite{ref:fractals}.

The Hofstadter  spectrum was originally obtained in a tight-binding model which describes  the motion on a square lattice of a (spinless) charged particle   hopping from one site to one of the nearest sites in the presence of an external magnetic field  perpendicular to the lattice plane. 

\begin{figure}[!thb]
\begin{center}
\includegraphics[width=3in]{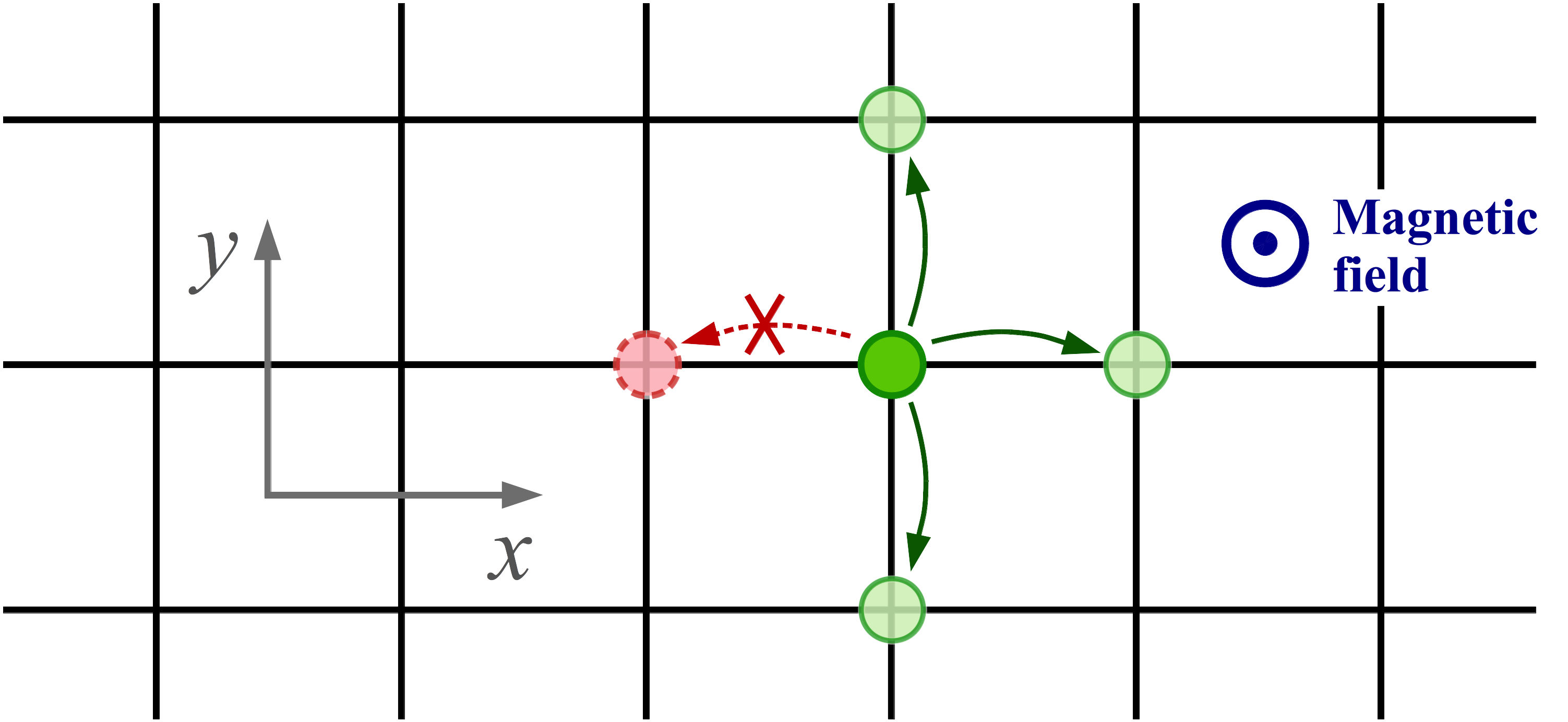}
\end{center}
\vskip -3mm
\caption{Biased motion in a magnetic field.}
\label{fig:setup}
\end{figure}

In this paper we address the problem of a {\emph{biased}} quantum motion where the motion is constrained -``biased''- by the condition that the hopping of the particle to the left is now forbidden, while it can still freely hop to the other nearest sites (right, up and down) as indicated in Fig.~\ref{fig:setup}. The particle is electrically charged so that its motion is affected by  the presence of the  magnetic field. Its dynamic  is described by a  truncated  non Hermitian Hofstadter  Hamiltonian.

This quantum system  was introduced in Ref.~\cite{ref:I} where its relation with biased classical random walks -- i.e. random walks on a square lattice 
conditioned 
to move horizontally only to the right, never to the left \cite{ref:Ibis} -- was studied. Explicit trace identities relating, on the one hand, the generating  functions of the algebraic area probability distribution   of  biased random walks to, on the other hand, the traces of powers of the quantum truncated  Hamiltonian, were proposed in the so-called commensurate case where the (normalised) magnetic flux per unit cell is a rational number. 
An exact solution for the quantum spectrum was derived. 
It turns out that the energy levels 
depend on two quantum numbers $k_x$ and $k_y$. The quasi-momentum $k_y$ takes continuous values in $[-\pi,\pi]$ and corresponds to the plane-wave Bloch states propagating along the vertical axis while the other quantum number $k_x$  takes discrete values in the same interval $ [-\pi,\pi]$ and fixes  the boundary condition on the horizontal axis. 

Following  the  approach of Ref.~\cite{ref:Hofstadter} where the fractal structure of the  Hofstadter band spectrum  is revealed when it is plotted as a function of the external magnetic flux, we are going to show below that the energy bands of the biased quantum model also exhibit, non surprisingly, a  self-similar structure in the same  ``energy-flux" plane. In addition, we will also argue that energy levels for different magnetic fluxes reveal -- when  they are plotted as a function of  $k_y$ -- unusual fractal patterns (``fractal carpets'') in the "energy-quasi-momentum" plane.

The  paper is organized as follows: in Sec.~\ref{eq:model} we consider the non-Hermitian Hofstadter model and its real-valued energy spectrum. In Sec.~\ref{eq:fractal} we argue  visually and numerically that the spectrum possesses a fractal structure  not only for the energy bands plotted against  the magnetic flux but also for the energy levels plotted against the quasi-momentum $k_y$. An analytical argument based on  Chebyshev nesting is presented in Sec.~\ref{sec:Chebyshev}. It is followed by yet another analytical argument for the flattened energy bands in Sec.~\ref{sec:flattened}. Complex-valued branches of the energy spectrum are briefly discussed in Sec.~\ref{sec:complex}. The last section is devoted to  our conclusions.

\section{The model and its spectrum}
\label{eq:model}

\subsection{The biased quantum model}

The biased quantum model  non-Hermitian  Hamiltonian is
\beqn
H_\beta = T_x + T_y + T^{-1}_y\,,
\label{eq:H:beta}
\eeqn
where the operators $T_x$ and $T_y$ describe the hopping of the  particle along (the positive directions of) the axes $x$ and $y$, respectively. Quantum mechanically, the hopping operators act on a state $\Psi_{m,n}$ at  lattice site $\iffalse{\boldsymbol{x}}= \fi (m,n)$ as ~\cite{ref:Zak}
\beqn
\begin{array}{rcl}
T_x \Psi_{m,n} & = & \Psi_{m+1,n}\,, \\[1.5mm]
T_y \Psi_{m,n} & = & e^{i 2 \pi m \beta} \Psi_{m,n+1}\,,
\end{array}
\label{eq:act}
\eeqn
where\footnote{Here we use the notations of Ref.~\cite{ref:Hofstadter}. In the notations of Ref.~\cite{ref:I} $\gamma = 2 \pi \beta$ has been used instead.}
\beqn
\beta = \frac{\Phi}{\Phi_0}\,, 
\label{eq:beta}
\eeqn
is the  flux $\Phi$ piercing an elementary plaquette of the square lattice counted in units of the elementary  flux quantum $\Phi_0 = h c/e$. The hopping operators obey the  commutation relation
\beqn
T_x T_y = e^{- 2 \pi i \beta} T_y T_x \,.
\label{eq:commutation}
\eeqn

The absence of the $T^{-1}_x$ operator in Eq.~\eq{eq:H:beta} -- corresponding to the absence of the hops to the left according to Fig.~\ref{fig:setup} -- makes the Hamiltonian $H_\beta$  non-Hermitian  because $T_{i}^\dagger \equiv T^{-1}_{i}$ for $i = x,y$ and, consequently, $H_\beta^\dagger \neq H_\beta$.

The normalised magnetic flux $\beta$ enters  only via Eqs.~\eq{eq:act} and \eq{eq:commutation}. Thus, the energy spectrum is periodic with respect to the integer shifts, $\beta \to \beta + 1$. By periodicity, we take $\beta$ in the interval $[0,1]$.  

\subsection{The energy spectrum}

\subsubsection{General solution}

In Ref.~\cite{ref:I} the spectrum of the  Hamiltonian~\eq{eq:H:beta} was derived in the commensurate case where the magnetic flux~\eq{eq:beta} is a rational number
\beqn
\beta = \frac{p}{q}\,
\label{eq:beta:rational}
\eeqn
with $p$ and $q$  naturally relative prime integers, so that their greatest common divisor is  equal to  unity
\beqn
{\mathrm{gcd}}(p,q) = 1\,.
\label{eq:gcd}
\eeqn

The  spectrum is obtained by solving 
\beqn
P_{p,q}(E,k_y) = e^{i q k_x}\,,
\label{eq:find:E}
\eeqn
where 
\beqn
P_{p,q}(E,k_y) = \prod_{r=1}^q \left[ E - 2 \cos \left( k_y + 2 \pi \frac{p}{q} r\right)\right]
\label{eq:Ppq}
\eeqn
is a degree $q$ polynomial of   the energy  $E$. Solutions of Eqs.~\eq{eq:find:E} and \eq{eq:Ppq} determine the possible energies
\iffalse
\beqn
E = E_{p,q} (k_x,k_y)\,.
\label{eq:E:sol:gen}
\eeqn
\fi
where the continuous variables $k_x \in [-\pi,\pi]$ and $k_y \in [-\pi,\pi]$ play the role of quasi-momenta and   the magnetic flux $\beta$ enters  in Eq.~\eq{eq:beta:rational} via the integers $p$ and $q$ which  label distinct  energy branches. 

The spectrum consists of the $q$ eigenenergies
\beqn
E^{}_{q,r}(k_y) = 2 \cos \left[\frac{1}{q} \left(\arccos\left[\cos(q k_y) + \frac{e^{i q k_x}}{2} \right] + 2 \pi r\right) \right]\!, \qquad
\label{eq:E}
\label{en} 
\eeqn
where $r = 1,2,\dots, q$ maps different branches of the solutions. 
It includes, in general, complex valued energies which correspond either to formal instabilities or to a dissipative motion of the particle. It turns out that the spectrum is real for at least certain intervals of the quasi-momentum  $k_y$  if and only if  $k_x$ is such that
\beqn
 e^{i q k_x} = \pm 1 \,,
\label{eq:s}
\eeqn
which corresponds to  periodic $e^{i q k_x} =1$ (antiperiodic $e^{i q k_x} =-1$) boundary conditions   along the $q$-site-long lattice cell in the $x$ direction. Below, we  restrict ourselves to periodic  and antiperiodic  boundary conditions  for which real eigenenergies do exist. 

An interesting feature of the spectrum is that it is independent of the integer  $p$ provided  $p$ and $q$ are relative prime integers as specified in Eq.~\eq{eq:gcd}.
\subsubsection{Examples and symmetries of the spectrum}

\begin{figure}[!thb]
\begin{center}
\includegraphics[width=3.2in]{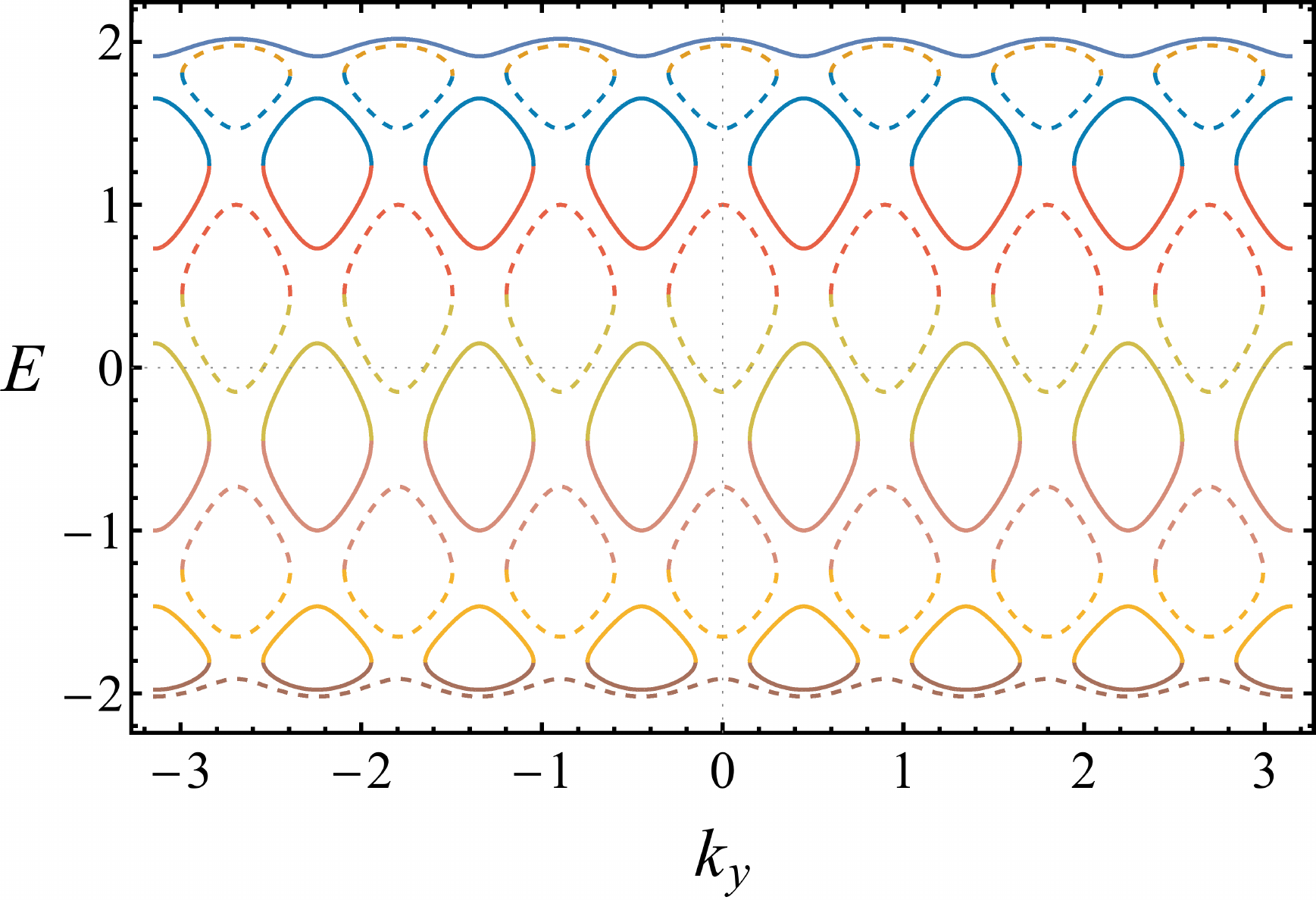}\\[1mm]
\includegraphics[width=3.2in]{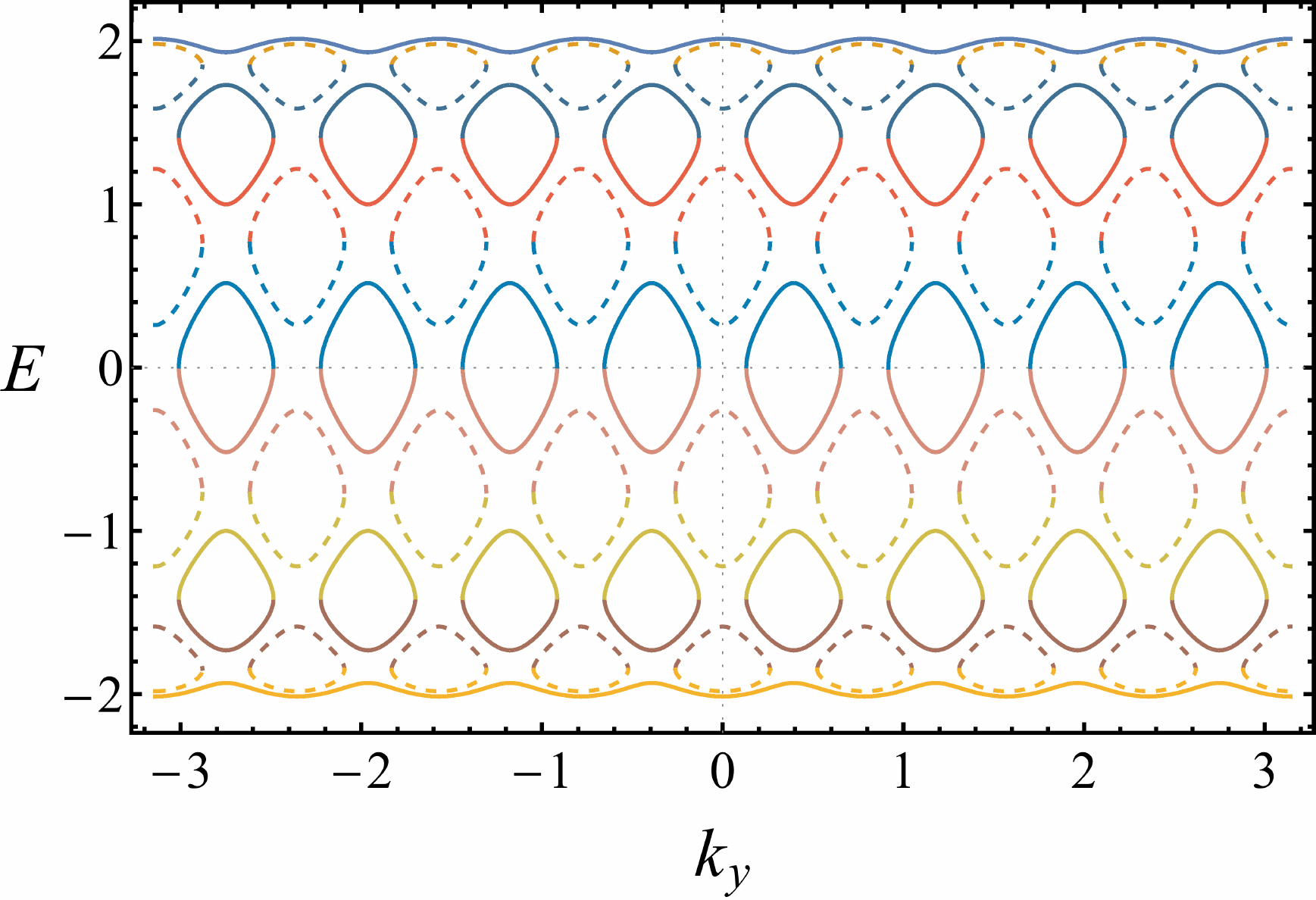}\\
\end{center}
\vskip -5mm
\caption{The energy spectrum $E_q$ for odd $q=7$ (top) and even $q=8$ (bottom) as the functions of the quasi-momentum $k_y$ for  periodic (solid lines) and antiperiodic (dashed lines) boundary conditions on the $x$ axis. The $q$ energy branches are marked by different colours.}
\label{fig:real:energy:q8}
\end{figure}

In Fig.~\ref{fig:real:energy:q8} (top) and (bottom) for $q=7$ and $8$ respectively,  the real energy levels are plotted as a function  of $k_y$ both for  periodic/antiperiodic boundary conditions (solid/dashed lines). One  sees that the $q$ real branches labelled by  $r = 1, 2, \dots q$ in Eq.~\eq{eq:E} do materialize for   particular intervals of $k_y$. 

 Since the features of Fig.~\ref{fig:real:energy:q8}  are generic for odd and even  $q$, we describe them in details below.

For odd $q$,   the highest/lowest real energy level is a continuous curve (solid/dashed)  for the whole quasi-momentum  interval $k_y \in [-\pi,\pi]$. The other $q-1$ branches are real only in $q$ distinct $k_y$ intervals  (here we use the fact that the spectrum lies on a circle thus the quasi-momenta  $k_y = - \pi$ and $ \pi$ are identified). These $q-1$ energy branches are grouped pairwise to form $(q-1)/2$ distorted ovals in each of these intervals. Thus, both in  the periodic and antiperiodic cases, one has one continuous energy band and $q (q-1)/2$ distorted energy ovals [see in Fig.~\ref{fig:real:energy:q8} (top) the  21 energy dashed/solid ovals for  $q=7$]. 

The odd-$q$ energy spectrum has a number of symmetries which are manifest  in Fig.~\ref{fig:real:energy:q8} (top):
\beqn
E^{(\pm)}_{q,r}(k_y) {\biggl{|}_{q \in \mathrm{odd}}} & = & E^{(\pm)}_{q,r}(- k_y) = E^{(\pm)}_{q,r}\left(k_y + \frac{2 \pi n}{q}\right) \nonumber\\
& = & - E^{(\mp)}_{q,r} \left(k_y + \frac{(2n+1) \pi }{q}\right)\,,
\eeqn
with $n \in \Z$. Indeed, the spectrum is symmetric under both mirroring in the quasi-momentum direction, $k_y \to -k_y$, and the discrete shifts $k_y \to k_y+ 2 \pi n/q$. Moreover, for odd $q$, the periodic and antiperiodic spectra  are related to each other by the mirroring transformation in the energy direction $E \to - E$ and a simultaneous shift along the quasi-momentum direction, $k_y \to k_y + \pi (2 n+1)/q$ (here we use again that in the quasi-momentum space  $k_y$ and $k_y + 2\pi$  are identified).

For even  $q$,  in the periodic case, both the highest and the lowest real energy levels are continuous curves for the whole interval $k_y \in [-\pi,\pi]$. The other $q-2$ branches are real only in $q$ distinct $k_y$ intervals. The $q-2$ energy branches are grouped pairwise to form $(q-2)/2$ distorted ovals in each of these intervals. Thus one has two continuous energy bands and $q(q-2)/2$ distorted energy ovals (see in Fig.~\ref{fig:real:energy:q8} (bottom) the  24 solid ovals for $q=8$). The antiperiodic case is a bit simpler since the $q$ energy branches are grouped pairwise to make $q/2$ separate energy ovals in the $q$ bands. Thus one has $q^2/2$  distorted energy ovals (see Fig.~\ref{fig:real:energy:q8} (bottom) the 32 dashed ovals for $q=8$). 

The symmetries of the even-$q$ spectrum are simpler:
\beqn
E^{(\pm)}_{q,r}(k_y) {\biggl{|}_{q \in \mathrm{even}}} & = & E^{(\pm)}_{q,r}(- k_y) = - E^{(\pm)}_{q,r} (k_y) \nonumber\\
& = & E^{(\pm)}_{q,r}\left(k_y + \frac{2 \pi n}{q}\right) \,,
\eeqn
with $n \in \Z$. The spectrum is symmetric with respect to mirroring of both energy $E \to -E$ and quasi-momentum $k_y \to - k_y$. The spectrum is also invariant under the discrete shifts of the momenta, $k_y \to k_y+ 2 \pi n/q$. Contrary to $q$ odd, for $q$ even the energy spectra for periodic and antiperiodic boundary conditions are not simply related.

Finally, due to the independence of the spectrum on the numerator $p$ of  the (normalised) rational magnetic flux $\beta$ in ~\eq{eq:beta}, the spectra in Fig.~\ref{fig:real:energy:q8} (top) (bottom) are realised, in fact, for six different values of  $\beta = p/7$ with $p = 1, \dots 6$.

\begin{figure}[!thb]
\begin{center}
\includegraphics[width=3.2in]{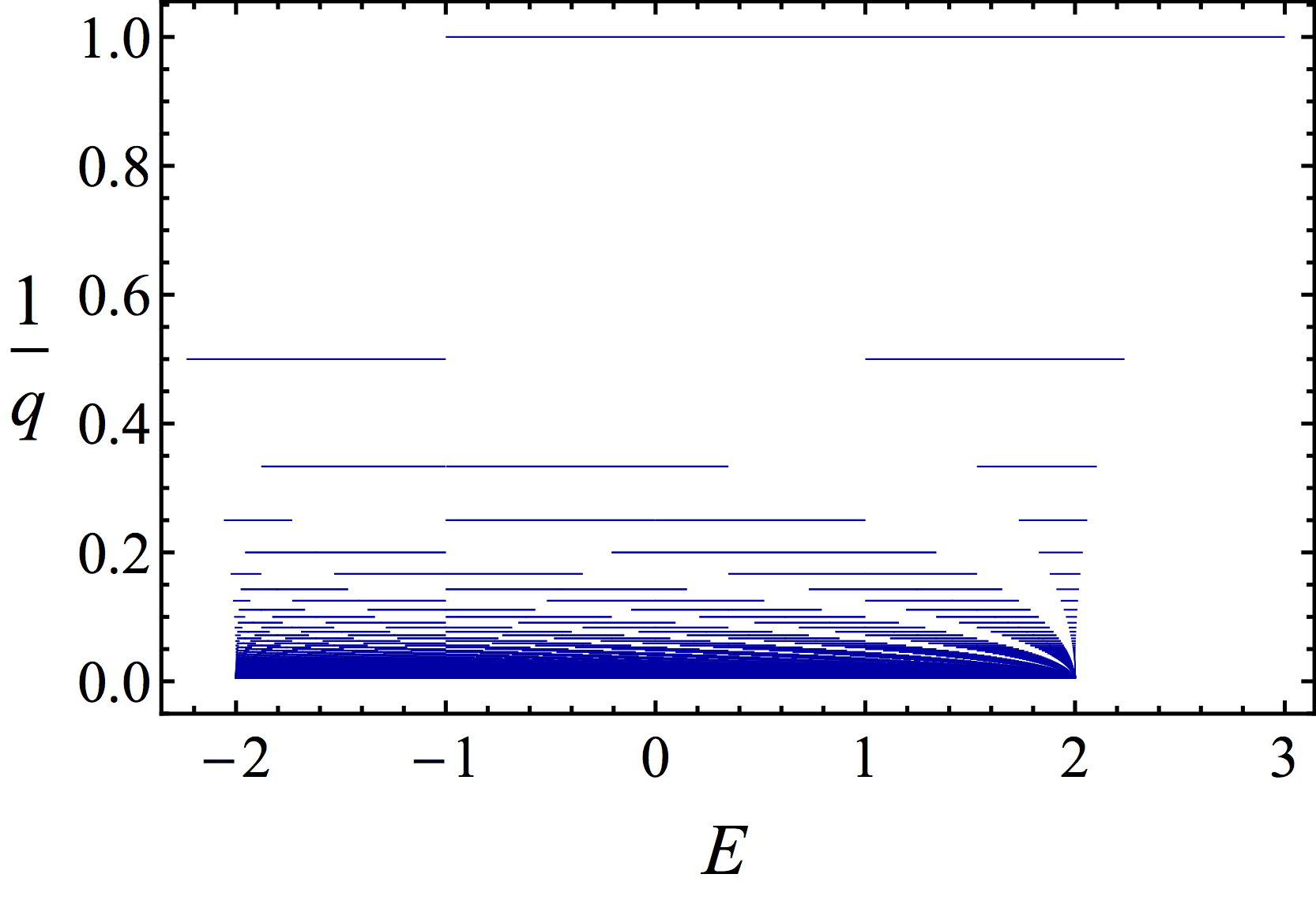}
\end{center}
\vskip -5mm
\caption{The band energy spectrum (the horizontal axis) with respect to the minimal value of the normalised magnetic flux $\beta^{\mathrm{}}=1/q\in[0,1]$ with $q=1,2,\ldots,100$ (the vertical axis) for periodic boundary conditions.}
\label{ref:projection}
\end{figure}

As clearly seen in Fig.~\ref{fig:real:energy:q8}, the real spectrum occupies the domain   $[-\pi,\pi]\otimes [-2,2]$ with energy bands     depending on the magnetic flux $\beta$. Following Ref.~\cite{ref:I} we plot in Fig.~\ref{ref:projection}  the energy bands   as a function of $\beta = 1/q$ for  $q=1,2,\ldots,100$. Notice that the band spectrum does not carry any information on the actual quasi-momentum dependence  since it is obtained by a 1d projection of the spectrum on the vertical (energy) axis. In the next section we are going to show that not only the $\beta$-dependence of the band spectrum but also the quasi-momentum $k_y$-dependence of the energy spectrum~\eq{eq:E} do have fractal self-similar patterns.

\section{Structure of energy levels}
\label{eq:fractal}

\subsection{Structure of energy bands}

It is well-known that the  Hofstadter band spectrum\footnote{Notice that in the biased quantum model the real-valued band spectrum lies in the narrower interval $E \in[-3,3]$.}  $E \in[-4,4]$ reveals a fascinating fractal structure -- the Hofstadter  butterfly -- when it is plotted on the horizontal axis against  the normalised magnetic flux $\beta \in [0,1]$ on the vertical axis. Likewise,  in the biased quantum model, a fractal structure  should also materialize when the energy bands are plotted against $\beta$. 

\begin{figure}[!thb]
\begin{center}
\includegraphics[width=3.3in]{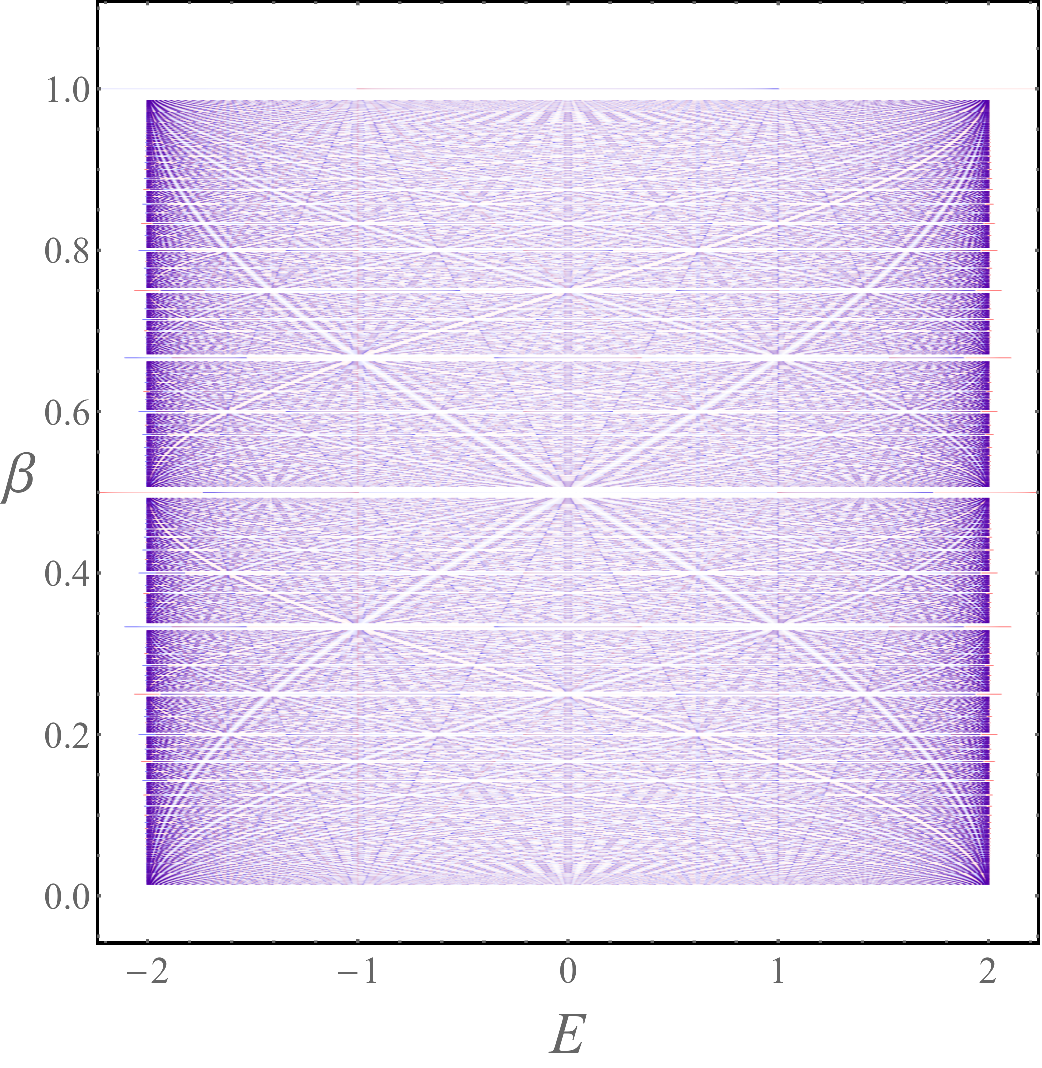}
\end{center}
\vskip -5mm
\caption{The band energy spectrum (the horizontal axis) with respect to the value of the normalised magnetic flux $\beta=p/q\in[0,1]$ with $q=1,2,\ldots,70$ and $p=1, \ldots, q$ (the vertical axis) for the periodic (the red lines) and antiperiodic (the blue lines) boundary conditions. The darkness/lightness of the lines visualise high/low values of the density of states~\eq{eq:DOS}.}
\label{fig:butterfly}
\end{figure}

In Fig.~\ref{fig:butterfly}  the band spectrum is plotted against $\beta$  with all  values  $ p/q \in [0,1]$ (and not only the minimal  values $ 1/q$ as  in Fig.~\ref{ref:projection}) for $q = 1,\ldots,70$,   both for periodic and antiperiodic boundary conditions. Here $q$ plays a role of a ``resolution'' parameter which determines the thinness of the chosen grid of possible $\beta$ values.

\begin{figure}[!thb]
\begin{center}
\includegraphics[width=3.2in]{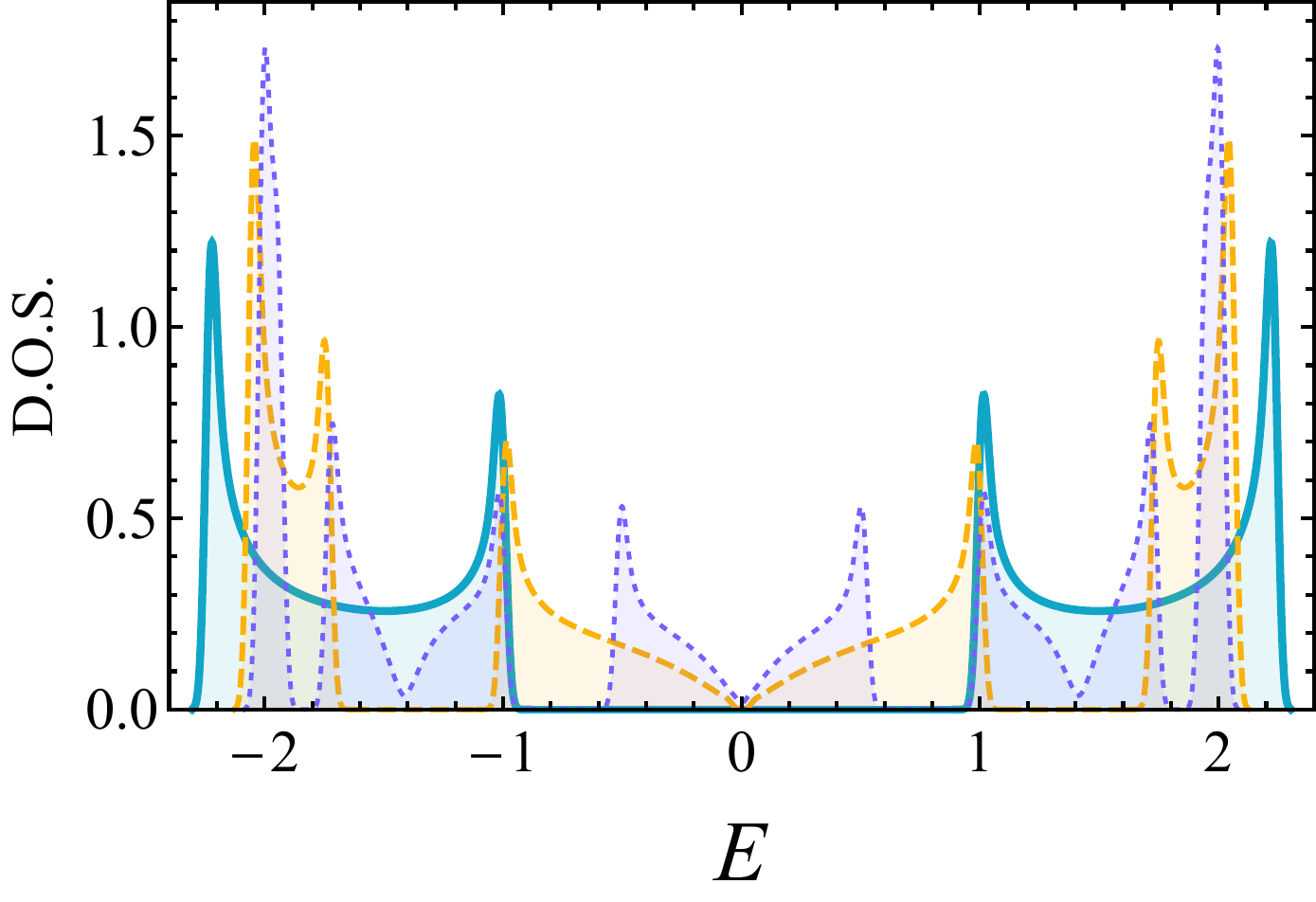} \\[3mm]
\includegraphics[width=3.2in]{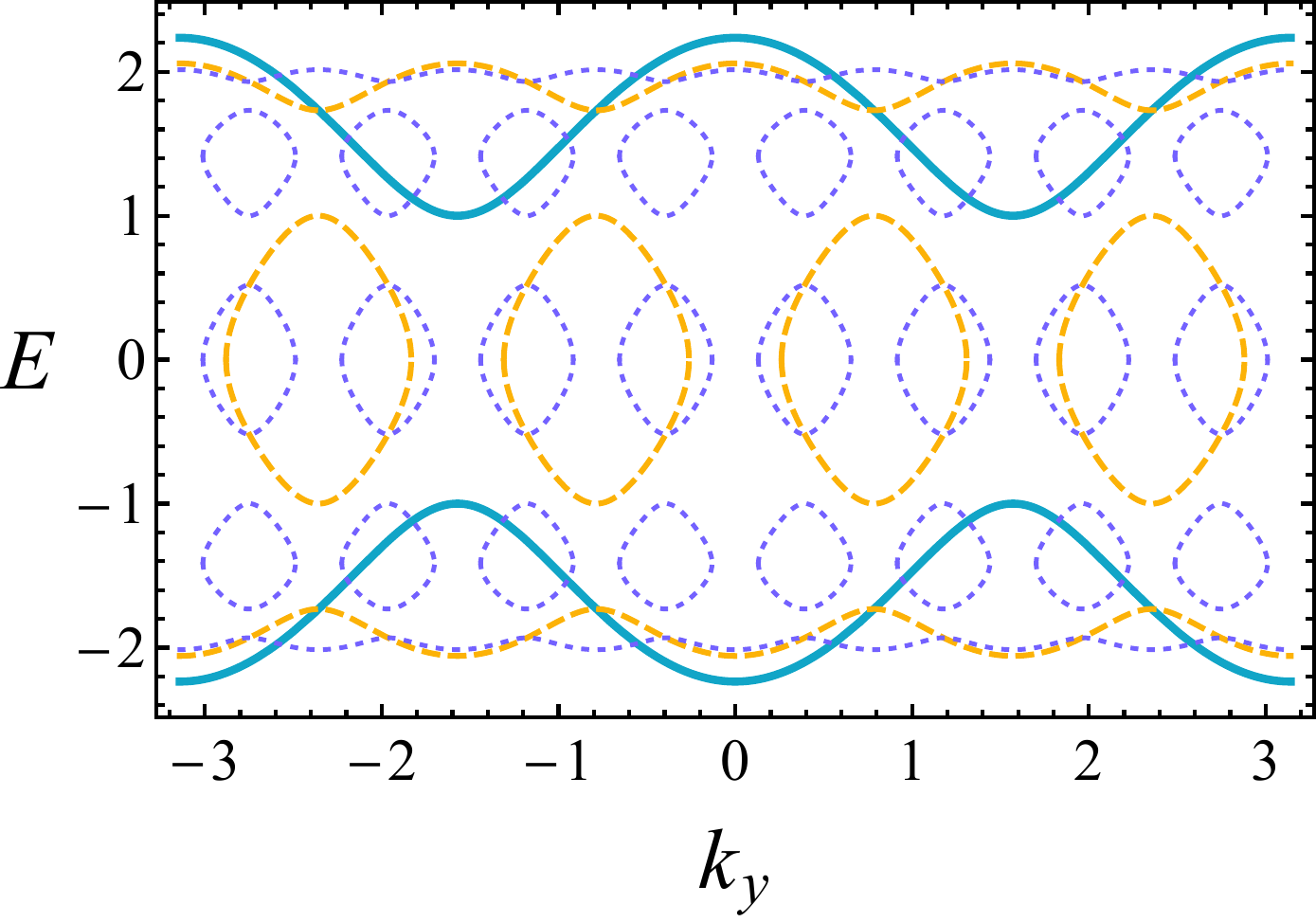}
\end{center}
\vskip -5mm
\caption{(top): the density of states~\eq{eq:DOS} as a function of the energy $E$ for $q=2$ (the blue solid line), $q=4$ (the orange long-dashed line) and $q=8$ (the violet short-dashed line) levels; (bottom): the  energy levels as a function of the quasi-momentum $k_y$.}
\label{fig:dos}
\end{figure}

In Fig.~\ref{fig:butterfly}  the variations of lightness in each energy band correspond to the variations of the (normalised) density of states at a given energy
\beqn
 \frac{1}{2\pi} \frac{d k_y}{d E}\,
\label{eq:DOS}
\eeqn
such that the higher/lower density of states, the darker/lighter the band.
As an illustration in Fig.~\ref{fig:dos} (top) and (bottom),  we display   the density of states~\eq{eq:DOS} and the corresponding  energy levels for $q=2, 4, 8$, respectively.

At a fixed resolution $q$ the (normalised) magnetic flux runs over the values
\beqn
\beta = 0, \ \frac{1}{q}, \ \frac{2}{q}, \ \dots, \frac{q-1}{q}\,,
\label{eq:beta:all}
\eeqn
where  $\beta = 0$ corresponds to $p = q$. The energy levels are defined by Eqs.~\eq{eq:find:E} and \eq{eq:Ppq} that is to say
\beqn
\prod_{r=1}^q \left[ E - 2 \cos \left( k_y + 2 \pi \beta r\right)\right] = e^{i q k_x}\,.
\label{eq:def:2}
\eeqn

Evidently, some of the instances of $p$ and $q$ in Eq.~\eq{eq:beta:all}  correspond to mutually reducible integers so that their greatest common divisor ${\mathrm{gcd}}(p,q) \neq 1$. In this case one can use the property of the energy polynomial~\eq{eq:Ppq} which appears in the LHS of Eq.~\eq{eq:def:2}
\beqn
\prod_{r=1}^q \left[ E - 2 \cos \left( k_y + 2 \pi \beta r\right)\right] {\biggl|}_{\beta = \frac{p}{q}} \!\! = Q^{{\mathrm{gcd}}(p,q)}_{q/{\mathrm{gcd}}(p,q)}(E,k_y), \qquad
\label{eq:rhs:Q}
\eeqn
where
\beqn
Q_q(E,k_y) = 2 \left[T_q\left(E/2\right) - \cos(q k_y)\right]\,,
\label{eq:Qq}
\eeqn
and $T_q(x)$ is a Chebyshev polynomial of the first kind\footnote{We use the standard notation $T_n$ for the Chebyshev polynomial of the first kind determined by the equation $T_n(\cos \varphi) = \cos(n\varphi)$.}.

Substituting Eq.~\eq{eq:rhs:Q} into Eq.~\eq{eq:def:2}, we get for the energy levels
\beqn
Q^{{\mathrm{gcd}}(p,q)}_{q/{\mathrm{gcd}}(p,q)}(E,k_y) = e^{i q k_x}\,,
\eeqn
which can be simplified as
\beqn
Q_{q_p}(E,k_y) = e^{i q_p k_x}\,,
\eeqn
where the integer  $q_p$ is
\beqn
q_p = \frac{q}{{\mathrm{gcd}}(p,q)}\,.
\label{eq:q:p}
\eeqn

We come to some trivial conclusions for the spectrum in the case of the rational (normalised) magnetic flux~\eq{eq:beta:rational} 
\begin{itemize}

\item[1.)] if $p$ and $q$ are mutual prime integers as in~\eq{eq:gcd}, then the energy spectrum is determined by  
\beqn
Q_{q}(E,k_y) = e^{iqk_x}\,,
\label{eq:Qqs}
\eeqn
because $q_1 \equiv q$.

\item[2.)] if $p$ and $q$ are not mutual primes [${\mathrm{gcd}}(p,q) \neq 1$], then the energy spectrum is determined by 
\beqn
Q_{q_p}(E,k_y) =  e^{iqk_x}\,,
\label{eq:Q:q:p}
\eeqn
with $q_p$ given in Eq.~\eq{eq:q:p}.

\item[3.)] in the special case $p=q$, where the magnetic field is absent, $\beta = 0$,  we  get the simple $q$-independent solution 
\beqn
E^{}_{1,1}(k_y)   =2 \left[\cos(k_y)+{ e^{i k_x}\over 2}\right]\,,
\label{eq:E0}
\eeqn
which is  Eq.~\eq{eq:E} for $q = 1$ and thus $r=1$.

\end{itemize}

For $p=1, 2, \dots q-1$ the integer $q_p$ runs over all possible divisors of the natural number $q$. Therefore the nontrivial energy levels at  resolution $q$ are determined by Eq.~\eq{eq:Q:q:p} where the $q_p$'s are taken from the set of all non equal divisors of  $q$.

For example, the scan of non zero possible normalised magnetic fluxes with resolution $q=6$ includes the series 
$\beta = 1/6, \ldots, 5/6$. According to our discussion, the mutually prime numbers in $\beta = 1/6, \, 5/6$ share the common spectrum corresponding to $\beta = 1/6$, while  $\beta = 2/6$ and $\beta = 4/6$ correspond to the spectrum of $\beta = 1/3$ and $\beta = 3/6$ corresponds to the spectrum of $\beta=1/2$. Therefore, the magnetic fluxes at resolution $q=6$ should include the solutions at $\beta = 1/2, 1/3, 1/6$ [plus, of course, the trivial solution~\eq{eq:E0}]. In other words, the solutions are given by Eq.~\eq{eq:E} with $q = 1,2,3,6$.

As yet another example, consider the magnetic fluxes with resolution $q=16$, given by $\beta = 1/16, \ldots, 15/16$, which can be divided into four different groups. The first group with the mutually prime numbers, $\beta = (2n + 1)/16$, $n = 0,1, \dots, 7$, has a spectrum corresponding to the minimal representative $\beta = 1/16$.  The second group, given by $\beta = 2 (2n + 1)/16$ with $n=0,\dots 3$, corresponds to the mutually prime numbers $\beta = (2 n +1)/8$ with the minimal representative $\beta = 1/8$. The third group is $\beta = 4/16, 12/16$ which has the same spectrum as $\beta = 1/4$. Finally, the last group includes only  $\beta = 8/16 = 1/2$. Thus, the magnetic fluxes at resolution $q=16$ should include the solutions corresponding to $\beta = 1/2, 1/4, 1/8, 1/16$ and~\eq{eq:E0}. In other words, the solutions are given by Eq.~\eq{eq:E} in which $q= 1,2,4,8,16$ runs over the set of all divisors of $16$.

Both examples illustrate that the energy levels with resolution $q$ are given by the the energy solutions~\eq{eq:E} in which $q_p$ runs over all possible divisors of the original resolution factor $q$ 
 [including, of course, the trivial energy level  Eq.~\eq{eq:E0}]. 
 
 The energy bands are plotted in Fig.~\ref{fig:butterfly}: they do have a characteristic pattern  which no
 longer  resembles a butterfly but rather  a  spider net.

\subsection{Fractal energy levels in quasi-momentum space}

\subsubsection{Self-similarity}

To address the fractal structure of the eigenenergies of the non Hermitian Hofstadter Hamiltonian~\eq{eq:H:beta}  plotted as a function of the quasi-momentum $k_y$ one has to consider curves made of all the disjoints energy ovals lying in the planar domain $[-\pi,\pi]\otimes [-3,3]$.  

Let us focus on the series $q=m^n$  where the base $m$ includes any integer $2, 3, 5, 6, 7, 10, \ldots$ which is not a member of another series with a smaller base (i.e. $m \neq m_1^{n_1}$ for any integers $m_1>1$ and $n_1 >1$). For example,  Fig.~\ref{fig:real:energy:q8} (top) for $q=7$ is the starting plot of the series $q=7^n$ with  base $m=7$ whereas  Fig.~\ref{fig:real:energy:q8} (bottom) for $q=8$ belongs to the $m=2$ series $q=2 \to 4 \to 8\to 16\to 32\to \ldots$.  We are going to argue that for a given base $m$  in the limit $n\to\infty$  the  curve made of the reunion of all the disjoint ovals in each of the $q=m^n$ members of the series constitute a fractal area--filling curve with Hausdorff dimension 2,  as it is the case for  well-known area--filling curves such as, for example, the Peano, Wunderlich or Hilbert curves~\cite{ref:fractals}.

\begin{figure}[tbph]
\begin{center}
\includegraphics[width=3.3in]{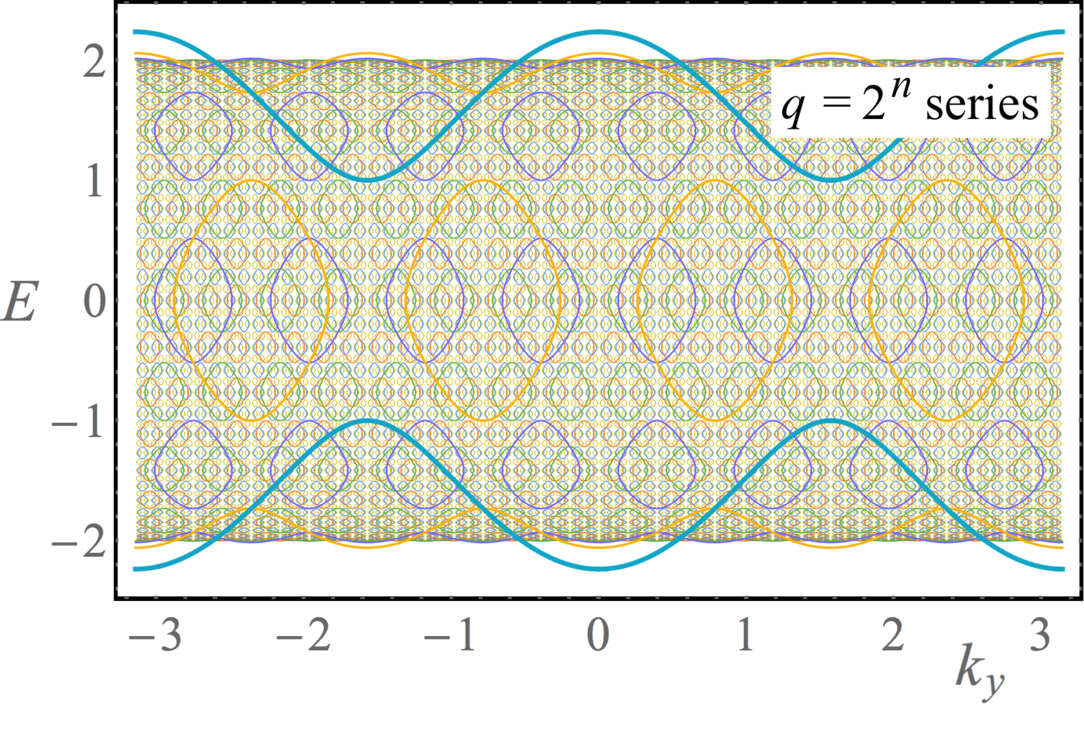}\\[-5mm]
(a) \\[3mm]
\includegraphics[width=3.3in]{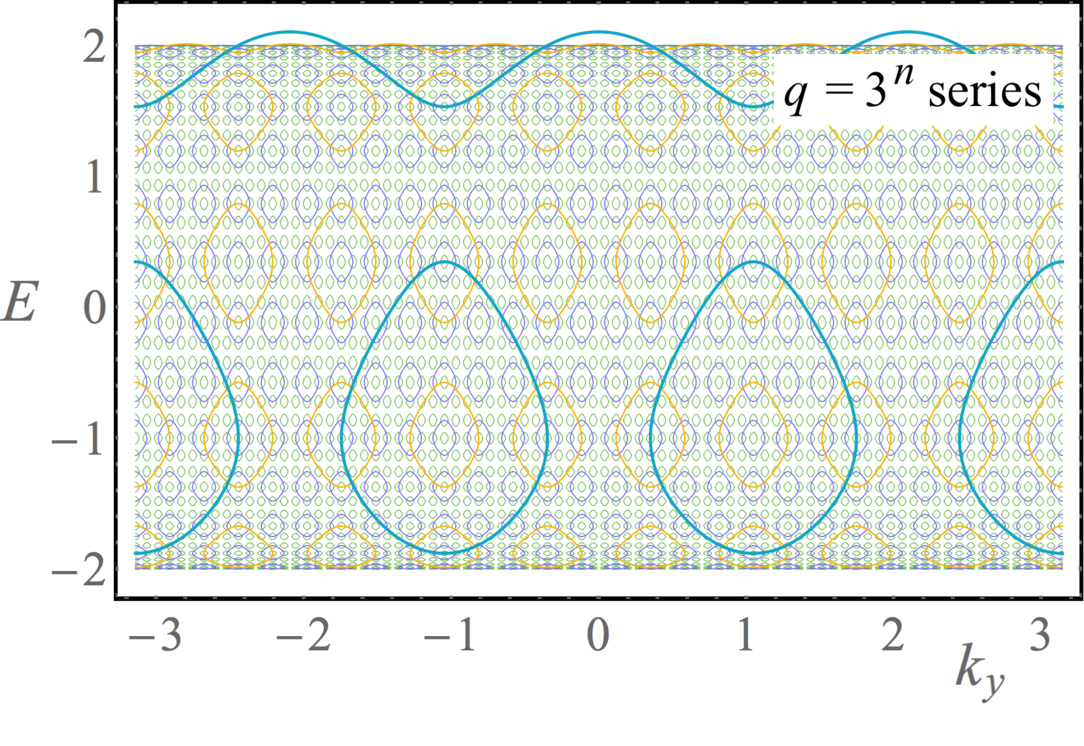}\\[-5mm]
(b) \\[3mm]
\includegraphics[width=3.3in]{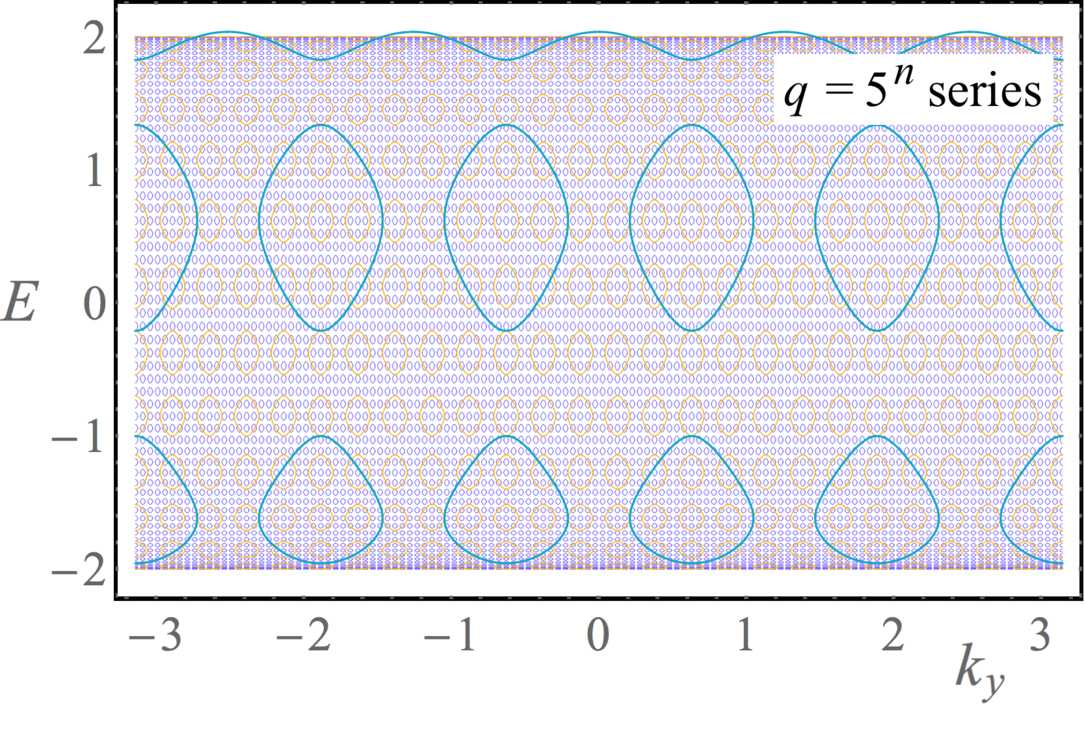}\\[-5mm]
(c)
\end{center}
\caption{The energy spectra~\eq{eq:E} are shown as functions of the quasi-momentum $k_y$ at (normalised) magnetic fields $\beta = 1/q$ for $q=m^n$ series with 
(a) $m=2$, $n = 1, \dots, 6$; 
(b) $m=3$, $n = 1, \dots, 4$; 
(c) $m=5$, $n = 1, \dots, 3$.}
\label{fig:2:3:5}
\end{figure}

\begin{figure}[!thb]
\begin{center}
\includegraphics[width=3.3in]{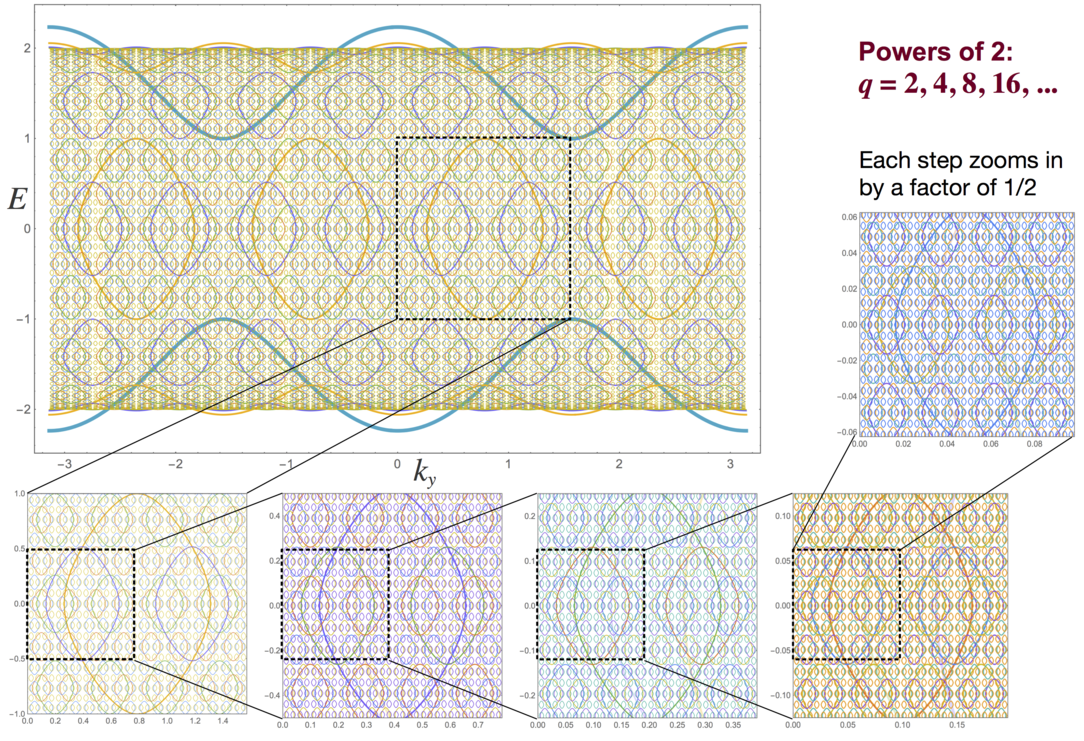}
\end{center}
\caption{Zooming into the fractal structure of the  energy levels of the non-Hermitian Hofstadter model for (normalised) magnetic fields $\beta = 1/q$ with $q = 2^n$ and increasing  $n$.}
\label{ref:zooming2}
\end{figure}

The curves for the series $q=m^n$ with  $m=2,3,5$ are shown in Fig.~\ref{fig:2:3:5}.
 Note that they  are not per se strictly  self similar because of the distortion of the ovals near the upper and  lower energies $E=\pm 2$. Leaving aside these spurious  boundary effects, the overall pattern is those of a fractal which remains unchanged by a  more an more accurate zooming. In the sequel we concentrate on the central part  of the energy plot, say  $E\in [-1,1]$, where  boundary effects become negligible and the self-similar structure of the oval patterns is  manifest. 
As an example, the  self--similarity of the energy levels in the $q=2^n$ series  is explicitly illustrated  in Fig.~\ref{ref:zooming2}.
It is clear that in the central region (far from the distorted boundaries) the $q=2^{n+1}$ energy levels are self-similar to a zoomed version of the $q=2^n$ levels, this for any power $n$. 
This self-similarity  depends on the base number $m$ of the $q = m^n$  series. Thus, in the limit $n \to \infty$, we have not one but rather an infinite number of fractals labeled by the base $m$ that one considers.

\subsubsection{Energy levels as space-filling curves}

There are several ways to determine the fractal structure of a 2d curve. One way is to consider a curve which, by successive iterations, would have a diverging length yet the area in which it is contained remains finite (typical examples of such curves are the Koch and Peano curves~\cite{ref:fractals}). The scaling of the length of the curve with the length's resolution should give us the Hausdorff -- fractal -- dimension of the curve. Let us first address this question  numerically and then, in the next section Section~\ref{sec:Chebyshev}, consider some more precise analytical arguments. 

As we have already seen, at resolution $q$ the number of energy ovals is $q(q-2)/2$ for an even $q$ and $q(q-1)/2$ for an odd $q$.
As a simple example,    let us again consider    the  $q=7$ and $q=8$ curves  of Figs.~\ref{fig:real:energy:q8}. Since the energy pattern repeats itself horizontally $q$ times, it is sufficient to consider the part of the curve made of the ovals contained in a given $k_y$ interval (a given column). 
Numerical estimates for $q=2^{n}, q=3^n, q=5^n$ and large $n$ (typically, the maximal power is $n=24$ for  base $m=2$) give for the  perimeter length ${\mathcal{L}}^{\mathrm{col}}$ and the  area  enclosed ${\mathcal{A}}^{\mathrm{col}}$ respectively 
 
\beqn
{\mathcal{L}}^{\mathrm{col}} =7.34771\,, \quad\quad 
{\mathcal{A}}^{\mathrm{col}}={7.74880\over q}\,.
\label{eq:LA:col}
\eeqn
As for the total perimeter length ${\mathcal{L}}^{\mathrm{tot}}$ and the total area ${\mathcal{A}}^{\mathrm{tot}}$ of the entire curve made of all the disjoint ovals, the scalings\footnote{Note that the area enclosed by the ovals is contained in a rectangle of area $2\pi\times 4\simeq 25.1327$ whereas the "available" surface inside the rectangle, i.e. after removing  the gaps between the bands, is~\cite{ref:I} in the large $q$ limit $2 \pi\times 8/3 = 16.7552$.} are respectively
\beqn
{\mathcal{L}}^{\mathrm{tot}} \simeq 7.34771 \times q\,, \quad\quad 
{\mathcal{A}}^{\mathrm{tot}}\simeq  7.74880\,.
\label{eq:LA:tot}
\eeqn

The  scaling~\eq{eq:LA:tot} -- a diverging length for  a curve which nevertheless encloses a finite area -- is typical of a fractal area--filling curve. The  Hausdorff dimension $D$  of the curve and the resolution of the curve's details   $1/q$ -- so that  the bigger the resolution $q$, the smaller the scale at which one looks at the curve (as  already mentioned,  $1/q$ corresponds to the minimal possible nonzero value of the normalised magnetic flux $\beta$) -- should be related by
\beqn
{\mathcal L}^{\mathrm {tot}} = \gamma {\left({1\over q} \right)}^{1-D}\,, \qquad {\mathcal A}^{\mathrm {tot}} = const\,.
\label{eq:LA:general}
\eeqn
A comparison of  Eq.~\eq{eq:LA:tot} with Eq.~\eq{eq:LA:general} gives a Hausdorff dimension $D=2$ which does correspond to an area-filling curve.

Another way to show that the Hausdorff dimension  is $2$ consists in rephrasing the scaling argument above in terms of iterative patterns which are clearly visible in Fig.~\ref{ref:zooming2} for  base $m=2$: each time the resolution is doubled, i.e. $1/q$ is divided by $2$, the number of ovals is multiplied by $4$. This property is again characteristic of a fractal with Hausdorff dimension $D= {\ln 4/ \ln 2}=2$.  In general for base $m$,  iterating means dividing $1/q$ by $m$ with a number of ovals multiplied by $m^2$, so again an Hausdorff dimension $D={\ln m^2/ \ln m}=2$.

\section{Chebyshev nesting and fractals}
\label{sec:Chebyshev}

The energy spectrum of the non-Hermitian Hamiltonian ~\eq{eq:H:beta} is determined by Eqs.~\eq{eq:Qq}, \eq{eq:Qqs} or \eq{eq:Q:q:p} which involve  Chebyshev polynomials of the first kind.  The fractal structure  of the  $q = m^n$ series can surprisingly be understood in terms of the so-called ``Chebyshev nesting'' which is a composition identity for  Chebyshev polynomials
\beqn
T_m\left(T_n(x)\right) = T_{mn}(x)\,.
\label{eq:nesting}
\eeqn

We can use three different approaches to discuss the relation of Chebyshev nesting~\eq{eq:nesting} to the fractal self-similarity of the energy curves.

Firstly, let us consider the horizontal scaling of the energy solutions of Eq.~\eq{eq:Qqs}.  As we have already noticed in Fig.~\ref{fig:real:energy:q8} and Fig.~\ref{fig:2:3:5}, the undistorted self-similar pattern of the energy levels is manifest in the vicinity of the $E=0$ horizontal axis. The same self-similarity should in particular  be seen in the scaling behaviour of the zero-energy solutions which occur for a set of specific quasi-momenta $k_y$. In other words, if, for  the series $q=m^n$, the energy curves exhibit a self-similarity  as the power $n$  increases, then one might  expect that the discrete quasi-momenta $k_y$ for which the energy lines cross the horizontal axis will also exhibit a self-similar scaling.  According to Eqs.~\eq{eq:Qq} and \eq{eq:Qqs} for periodic boundary conditions -- where indeed energy lines do cross the horizontal axis for both even and odd $q$ -- these zero-energy quasi-momenta are given by
\beqn
T_{m^n}(0) = \cos\left(m^n k_y\right)+ \frac{1}{2} \,.
\label{eq:T:sol:periodic}
\eeqn

Using the Chebyshev nesting~\eq{eq:nesting} as well as 
\beqn
T_q(0) = \cos\frac{\pi q}{2} \equiv
\left\{\begin{array}{rlr}
0 & \mbox{\ for odd $q$;} & \\
+1 & \mbox{\ for even $q$,} & \mbox{even $q/2$;}\\
-1 & \mbox{\ for even $q$,} & \mbox{odd $q/2$;}\\
\end{array}
\right. \quad
\eeqn
and
\beqn
T_q(+1) = +1\,, \qquad T_q(-1) = (-1)^q\,,
\eeqn
one arrives at
\beqn
T_{m^n} (0) {\biggl{\rvert}}_{n \geqslant 2} \equiv \underbrace{T_m(T_m(\dots T_m(0)))}_{\text{$n \geqslant 2$ times}} = f^+_m\,,
\label{eq:Tmn0}
\eeqn
where
\beqn
f^\pm_m = \frac{1 \pm (-1)^m}{2}
\equiv \left\{
\begin{array}{ll}
1 & \ \mbox{even/odd $m$},\\
0 & \ \mbox{otherwise}.
\end{array}
\right.
\label{eq:f}
\eeqn

The important consequence of~\eq{eq:nesting} is that, according to Eq.~\eq{eq:Tmn0},  $T_{m^n} (0)$ does not depend on $n$ for  $n \geqslant 2$. Then, in the periodic case at hand, the zero-energy quasi-momenta solutions of Eq.~\eq{eq:T:sol:periodic} are 
\beqn
k^{(m,n,l)}_y(E=0) = \frac{\pi}{6} \frac{1}{m^n} \left[3 - (-1)^m + 12 l\right]\,, 
\label{eq:k0:periodic}
\eeqn
where $l \in \Z$ labels the possible independent solutions contained in the interval $[-\pi,\pi]$. 

According to Eq.~\eq{eq:k0:periodic}, for any fixed two powers $n$ and $n+p$ with $p \in \Z$, the zero-energy quasi-momenta  are related by
\beqn
k^{(m,n+p,l)}_y(E=0) = \frac{1}{m^p} k^{(m,n,l)}_y(E=0)\,,
\label{eq:k:scaling}
\eeqn
This  implies that the zero energy quasi-momenta for $q=m^{n+1}$ can be obtained by a rescaling of the $q=m^n$ quasi-momenta by a factor $1/m$. In other words, the width of the ovals diminishes by $1/m$ with each step $n \to n + 1$. 

Secondly, let us consider the vertical scaling of the energy ovals for  periodic boundary conditions. This scaling can be determined in a similar way using a simple argument based on the Taylor expansion of  Chebyshev polynomials. Indeed, let us determine from Eqs.~\eq{eq:Qq} and \eq{eq:Qqs} the spectrum  in the region  close to  the $E=0$ axis  where the energy spectrum is undistorted.  For a given $q=m^n$, the Chebyshev polynomial in Eq.~\eq{eq:Qq} can be expanded in power series as
\beqn
T_{m^n} \left(x\right) & = & g_{m,n}(m^n x) + O\bigl(x^3\bigr)\,, 
\label{eq:Tmn:small}
\eeqn
where
\beqn
g_{m,n}(y) = f^+_m + f^-_m \left[f^+_{\frac{m-1}{2}} + (-1)^n f^-_{\frac{m-1}{2}}\right] y - \frac{f^+_m}{2}  y^2. \qquad  
\label{eq:gmn}
\eeqn

The first two terms in the small-$x$ expansion \eq{eq:Tmn:small}  depend solely on  $m^n x$. The next--order term $x^3$ in Eq.~\eq{eq:Tmn:small} does not share the same scaling, but in the small energies region (or, equivalently, small $x$), it can be neglected as well as the higher--order terms. These facts play an important role for the fractal properties of the energy spectrum: indeed keeping only these lowest--order terms,
  $g_{m,n}$ is invariant with respect to the shift $n \to n + 2$
\beqn
g_{m,n+2}(y) = g_{m,n}(y)\,
\eeqn
One can then distinguish two sectors  according to the  parity of $n$ 
\beqn
g^{\mathrm{even}}_m(y) \equiv g_{m,n}(y)\biggl{\rvert}_{n \in \text{even}} & = & f^+_m + f^-_m y - \frac{f^+_m}{2}  y^2, \qquad 
\label{eq:gmn:even}\\
g^{\mathrm{odd}}_m(y) \equiv g_{m,n}(y)\biggl{\rvert}_{n \in \text{odd}} & = & f^+_m + y \sin\frac{\pi m}{2} - \frac{f^+_m}{2}  y^2, \qquad 
\label{eq:gmn:odd}
\eeqn
where $\ell_n = \mathrm{even}\ \mathrm{(odd)}$ for even (odd) $n$, respectively. For  a given  $\ell_n$, $g^{\ell_n}_{m}$  is independent on ~$n$. 

Clearly, for  $q=m^n$ with periodic boundary conditions, the energy  close to the horizontal axis   is determined by 
\beqn
g^{\ell_n}_{m}\left(\frac{m^n E}{2}\right) =  \cos (m^n k_y)+\frac{1}{2} \,.
\label{eq:g:ell}
\eeqn
 Due to the independence on $n$ of the left hand side of Eq.~\eq{eq:g:ell} once the parity $\ell_n$ is fixed, the energy solutions for $q=m^{n+2p}$ and $q = m^n$ are related by
\beqn
E^{(m,n+2 p)} (k_y) & = & m^{-2 p} E^{(m,n)}(m^{2 p} k_y)\,,
\label{eq:E:k:scaling}
\eeqn
where $p \in \N$ is an arbitrary natural number. The self-similar scaling property~\eq{eq:E:k:scaling} is consistent with the zero energy quasi-momenta scaling~\eq{eq:k:scaling}.

The explicit solutions of Eqs.~\eq{eq:gmn:even}, \eq{eq:gmn:odd}, \eq{eq:g:ell},
\beqn
E^{(m,n)} = \left\{
\begin{array}{lll}
\pm\!\! & \frac{\sqrt{1 - 2 \cos\left(m^n k_y\right)}}{m^n}  & \quad \mbox{even $m$}\,, \\[3mm]
& \frac{1+2 \cos\left(m^n k_y\right) }{m^n} {(-1)}^{\frac{(m-1)n}{2}} & \quad \mbox{odd $m$}\,,
\end{array}
\right.\quad
\eeqn
are  low-energy solutions  valid  in the region $m^n |E| \ll \pi$. They highlight the self-similarity properties~\eq{eq:k:scaling} and \eq{eq:E:k:scaling} of the energy spectrum as  $n$ increases.

Thirdly, there is a bit more elegant and general way to show the fractal self-similarity of the spectrum both for periodic and antiperiodic boundary conditions by using yet another property of the Chebyshev polynomials
\beqn
\lim_{n\to \infty} \left[ (-1)^{m n} T_{m^n}(m^{-n} x)\right] = \left\{
\begin{array}{lr}
\cos x, & \ \mbox{even $m$}, \\[2mm]
\sin x, & \ \mbox{odd $m$}. \\
\end{array}
\right.
\quad
\label{eq:Tmn:limit}
\eeqn

For practical purposes it is convenient to use an approximate formula based on~\eq{eq:Tmn:limit}
\beqn
T_{m^n}(m^{-n} x) \to T^{(0)}_{m^n}(m^{-n} x) & = & f_m^+ \cos x 
\label{eq:Tmn:approximation}\\
& & + (-1)^{mn} f_m^- \sin x\,, \nonumber 
\eeqn
where $f_m^\pm$ is given in Eq.~\eq{eq:f}. The approximation made in~\eq{eq:Tmn:approximation} is valid in the interval $x\in [-\pi,\pi]$ with an error $10^{-4}$ for $m=2, n=4$. The accuracy of this approximation is increasing rapidly as either $m$ or $n$ or both become larger (for example for $m=n=4$ the error is less than $10^{-8}$).

For large $n$   one can combine, both for periodic and antiperiodic boundary conditions, Eqs.~\eq{eq:Qq}, \eq{eq:Qqs} and \eq{eq:Tmn:approximation} into 
\beqn
\begin{array}{lrcl}
\mbox{even $m$:} \quad & \cos \frac{m^n E}{2}  & = &  \cos\left(m^n k_y\right)\pm{1\over 2} \,, \\[3mm]
\mbox{odd $m$:} \quad & (-1)^{n} \sin \frac{m^n E}{2} & = &  \cos\left(m^n k_y\right)\pm{1\over 2} \,,
\end{array} \qquad
\label{eq:master:s}
\eeqn
which are again valid in  the energy domain $m^n |E| \ll \pi $. One can check that the solutions of Eqs.~\eq{eq:master:s} reproduce the lowest  energy levels. 

The symmetries of Eqs.~\eq{eq:master:s} clearly imply the fractal self-similarity  of the energy levels 
\beqn
E^{(m,n+p)} (k_y) = (-1)^{m p} m^{-p}  E^{(m,n)}\left(m^p k_y\right)\,,
\label{eq:E:scaling:full}
\eeqn
both for even and odd  $m$ and $p \in \N$. These equations are  consistent with the scalings \eq{eq:k:scaling} and \eq{eq:E:k:scaling}.

Equation~\eq{eq:E:scaling:full} implies that the energy spectrum for $q=m^{n+1}$ can be obtained by
\begin{enumerate}
\item[1.)] rescaling (i.e., squeezing) the $q=m^n$ solution by the factor $1/m$ both in the energy $E$ and quasi-momentum $k_y$ coordinates, and

\item[2.)] periodically copying (i.e., extending) the squeezed solution $m$ times along the energy $E$ axis and $m$ times along the quasi-momentum $k_y$ axis.
\end{enumerate}

Note that these properties cannot be readily seen from the original Eq.~\eq{eq:E}.
They highlight   three important properties of the $q=m^n$ series which were already numerically illustrated above. As   $n\to n+1$:
\begin{itemize}

\item[A)] the total area ${\cal A}_n$ enclosed by the energy ovals in a fixed {\it area} of the $(k_y,E)$ plane remains constant since the ovals become $m$--times smaller in each of the two ($k_y$ and $E$) directions while their number increases by a factor $m^2$  so that ${\cal A}_{n+1} = {\cal A}_n$;

\item[B)] the total lengths $\cL^{\mathrm{col}}_n$ and $\cL^{\mathrm{row}}_n$ of the energy oval perimeters in a fixed {\it column} and a fixed {\it row}  respectively remain constant since they become $m$--times smaller  while their numbers    in each column and in each row increase by a factor $m$ so that 
$\cL^{\mathrm{col}}_{n+1} = \cL^{\mathrm{col}}_{n}$ and $\cL^{\mathrm{row}}_{n+1} = \cL^{\mathrm{row}}_{n}$;

\item[C)] the total length of the energy oval perimeters ${\cal L}^{\mathrm{tot}}_n$ in a fixed {\it area} in the $(k_y,E)$ plane increases by a factor $m$ so that ${\cal L}^{\mathrm{tot}}_{n+1} = m {\cal L}^{\mathrm{tot}}_{n}$ due to the second property  as well as because  the number of columns and ovals increases by the same factor $m$.

\end{itemize}

When $n \to \infty$  these  three properties  imply that the total length of the energy curves (i.e., of the oval perimeters) in a fixed area of the $(k_y,E)$ plane diverges as ${\cal L}^{\mathrm{tot}}_n \propto m^n$ while the area ${\cal A}$ enclosed by these ovals remains constant. Thus, a union of the energy ovals indeed constitute an area-filling curve with  Hausdorff dimension $D=2$. In agreement with Eq.~\eq{eq:LA:general}, the size $a_n \propto 1/q \equiv 1/m^n$ of any individual oval along any direction decreases with an increasing  resolution, $q\to \infty$, so that one gets ${\cal L}^{\mathrm{tot}}_n \propto a_n^{-1} \equiv a_n^{1-D}$ with the usual  $D=2$ Hausdorff dimension for space-filling curves.

Notice   that the band structure of the energy levels  has been obtained by projecting the $k_y$--dependent energy levels on the vertical energy axis and plotting it against the the magnetic flux $\beta$. Since the quasi-momentum dependent energy levels are self-similar, we conclude that for a fixed base $m$ the energy band spectrum has also to exhibit a fractal self-similarity for the different powers $n$ of the normalised magnetic flux $\beta = 1/m^n$, as illustrated in Fig.~\ref{fig:butterfly}.

\section{Flattened energy spectra}
\label{sec:flattened}

\subsection{Flattening}

The fractal self-similarity of the real-valued energy spectrum~\eq{eq:E} can  be addressed in yet another  way. As stressed above, the energy ovals  are  distorted near both edges of the spectrum $E \sim \pm 2$ (as illustrated in Fig.~\ref{fig:2:3:5}). On the other hand in the center $E \sim 0$, the distortion is practically absent and the self-similarity is quite pronounced. It turns out that in order to remove the boundary distortion effects from the entire energy domain, it is sufficient to consider, instead of the energy itself, the  energy-related quantity
\beqn
\cE^{}_{q,r}(k_y) = \arccos \left[\frac{1}{2} E^{}_{q,r}(k_y)\right]\,
\label{eq:arccos:E}
\eeqn
in the interval  $\cE\in[0,\pi]$.

Using ~\eq{eq:E} for  the $q$ energy branches $r = 1, \dots, q$ of the original  spectrum, one can rewrite Eq.~\eq{eq:arccos:E}
\beqn
\cE^{}_{q,r}(k_y) {=} f \left[\frac{1}{q} \left(\arccos\left[\cos(q k_y) + \frac{e^{iqk_x}}{2} \right] + 2 \pi r\right) \right] \qquad
\label{eq:E:temp}
\eeqn
where 
\beqn
f(x){\biggl\vert}_{x \in [0,2 \pi]} = \pi - |x - \pi| \equiv \arccos (\cos x)\,.
\eeqn

\begin{figure}[!htb]
\begin{center}
\includegraphics[width=3.3in]{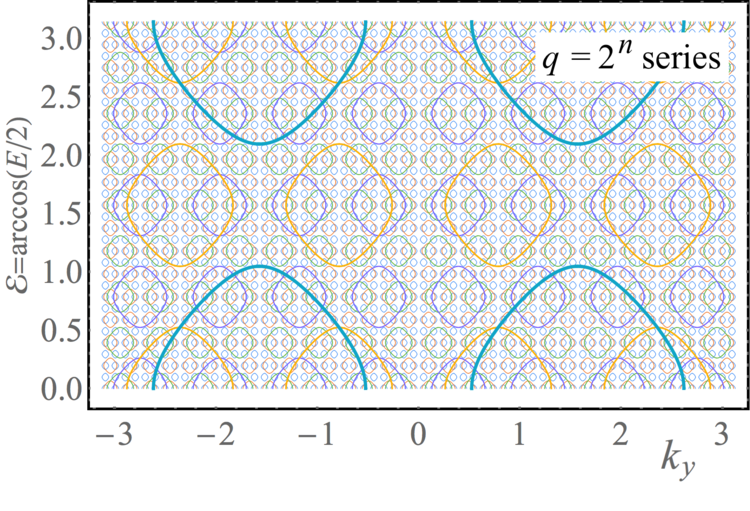}\\[-5mm]
(a) \\[3mm]
\includegraphics[width=3.3in]{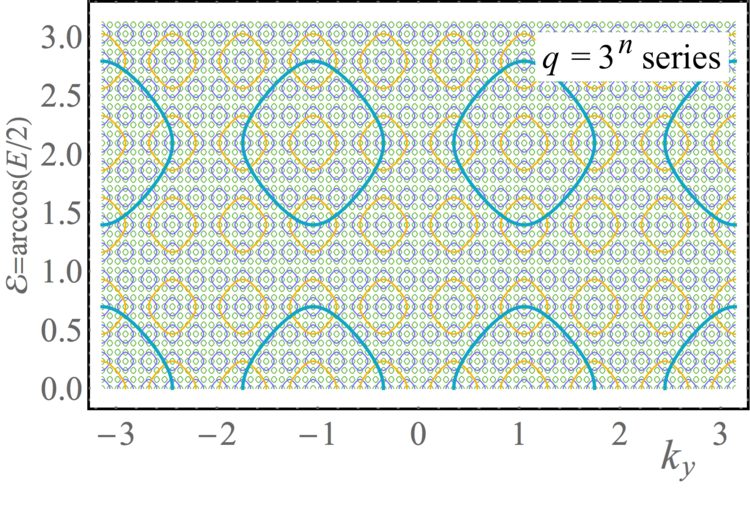}\\[-5mm]
(b)\\[3mm]
\includegraphics[width=3.3in]{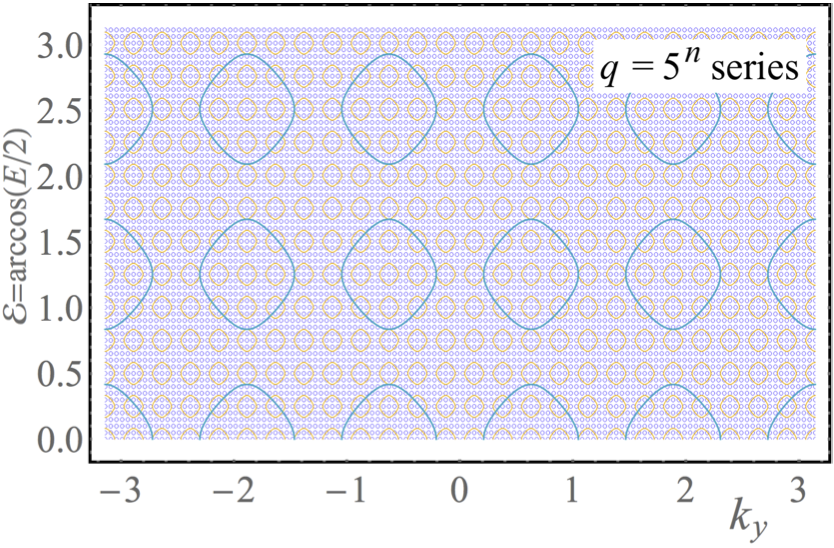}\\[-5mm]
(c)
\end{center}
\caption{The flattened energy spectra~\eq{eq:arccos:E} vs. the quasi-momentum $k_y$ for $q=m^n$ series with (a) $m=2$, $n = 1, \dots, 6$,  (b) $m=3$, $n = 1, \dots, 4$, and (c) $m=5$, $n = 1, \dots, 3$. These plots correspond the original energy spectra shown in Fig.~\ref{fig:2:3:5}.}
\label{fig:flattened:2:3:5}
\end{figure}

The flattened spectra~\eq{eq:arccos:E} are shown in Fig.~\ref{fig:flattened:2:3:5}(a), (b) and (c), respectively for the $q=m^n$ series with  $m=2$, $3$ and $5$ -- they correspond to the unflattened spectra of Fig.~\ref{fig:2:3:5}. Clearly, the arccos-operator in Eq.~\eq{eq:arccos:E} ``unfolds'' the  spectrum by flattening the energy levels close to  $E =\pm 2$, so that no distortion is  present anymore. It follows that the fractal self-similarity of the spectrum is manifest in the whole energy domain.

 For example in Fig.~\ref{fig:flattened:2:3:5}(a), each of the $q=2^{n}$ energy ovals intersects with two smaller $q=2^{n+1}$ energy ovals for  $n= 1, \dots 5$. Analogously in Fig.~\ref{fig:flattened:2:3:5}(b), each of the $q=3^{n}$ energy ovals intersects with four and includes  one smaller $q=3^{n+1}$ energy ovals for  $n= 1, \dots 3$. And  in Fig.~\ref{fig:flattened:2:3:5}(c) a similar self-similarity pattern can easily be observed for the  $q=5^n$ series.

\subsection{Self-similarity for the flattened bands}

\begin{figure}[!htb]
\begin{center}
\includegraphics[width=3.2in]{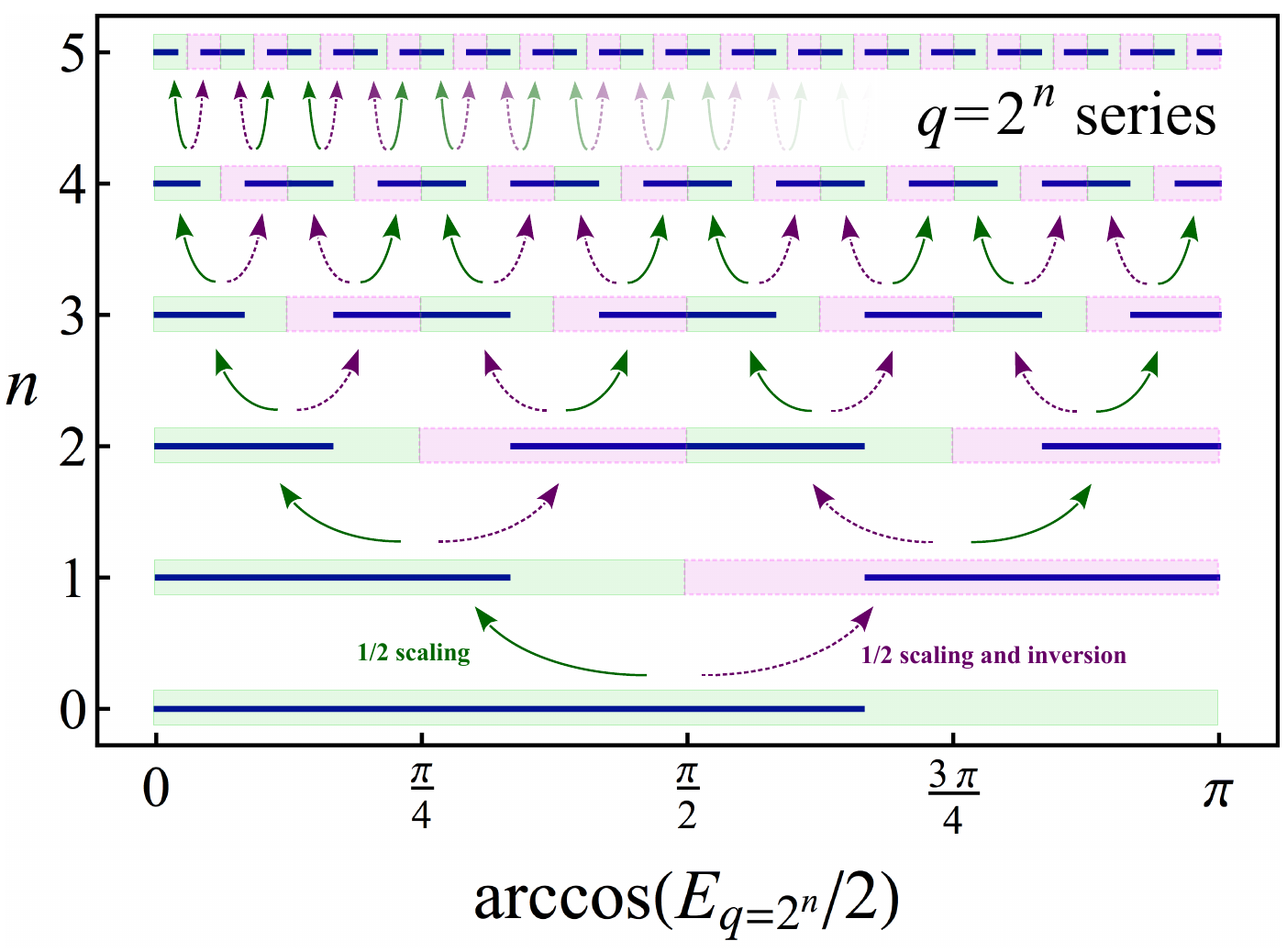}\\
(a) \\[6mm]
\includegraphics[width=3.2in]{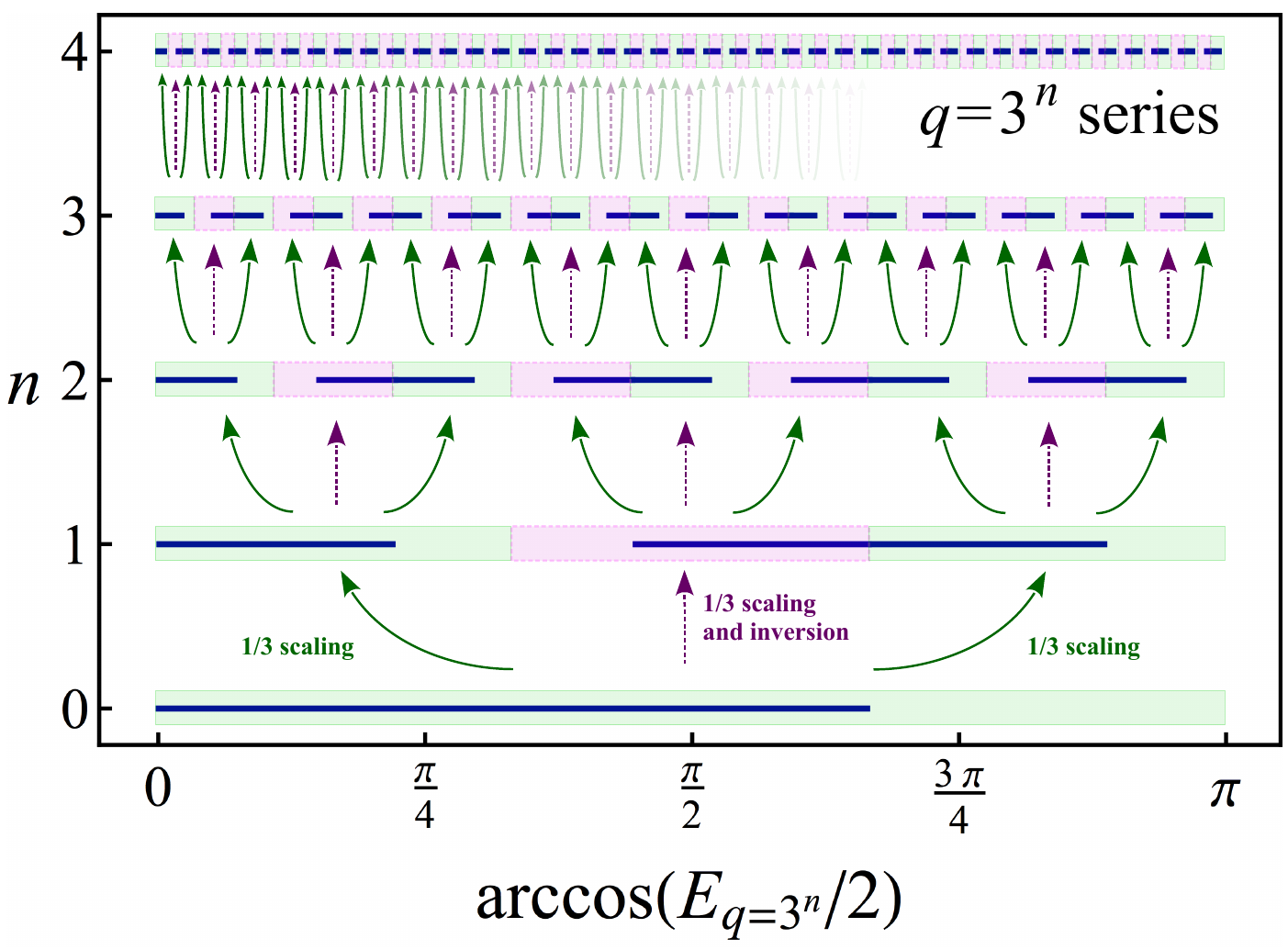}\\
(b) \\[6mm]
\includegraphics[width=3.2in]{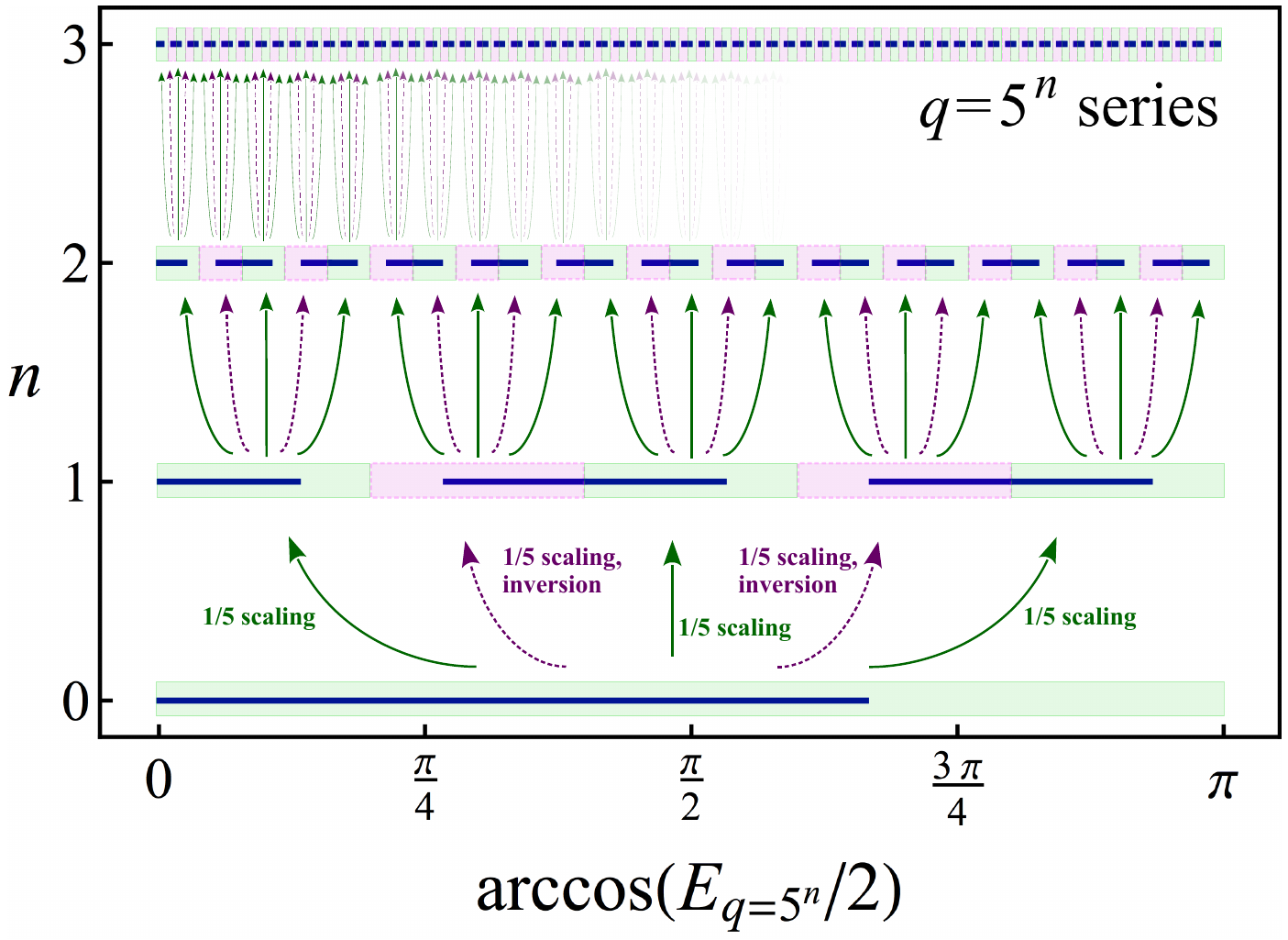}\\
(c)
\end{center}
\caption{The blue horizontal lines: The arccosine~\eq{eq:arccos:E} of the energy levels~\eq{eq:E} for the series $q=m^n$ at fixed power $n$ and for (a) $m=2$, (b) $m=3$ and (c) $m=5$. The shadowed regions and the arrows show the self-similarity features.}
\label{fig:self-similarity}
\end{figure}

Besides these visual observations, one can make a more precise  statement on the self-similarity. In Fig.~\ref{fig:self-similarity} we plot the flattened band spectra for $q=m^{n}$ with $m=2,3,5$. These plots share certain features with Fig.~\ref{ref:projection} but with the difference that in Fig.~\ref{fig:self-similarity} the vertical axis shows the power $n$ for a given base $m$ instead of $1/q$ for a set of $q$'s.

Consider first  in Fig.~\ref{fig:self-similarity}(a) the  $q=2^n$ series. The spectrum at the lowest level $n=0$ has -- in terms of the flattened energy variable~\eq{eq:arccos:E} -- a single band of  width $\delta\cE_{2^0}=2 \pi/3$ and a gap of width $\pi/3$. In order to get the spectrum at the next level $n=1$, it is sufficient to 
\begin{itemize}
\item[(i)] rescale the original $n=0$ spectrum, shrinking it by a factor $1/2$ so that $\cE\in[0,\pi]\to \cE\in [0,\pi/2]$

\item[(ii)] duplicate the resulting spectrum to cover again the whole energy range, $\cE\in[0,\pi/2]\cup[\pi/2,\pi]\to \cE\in [0,\pi]$

  \item[(iii)] and finally invert the duplicate around its center $\cE = 3\pi/4$.
  \end{itemize}
Obviously,  the total width of the two resulting bands at  level $n=1$  remains equal to the width of the single original $n=0$ band, $\delta\cE^{\mathrm{tot}}_{2^1}  = \delta\cE_{2^0}$.

Recursively, the band spectrum   at level $n+1$ is obtained  by applying the same rescaling/duplicating/inverting procedure on the spectrum at level $n$:
\begin{itemize}
\item[(i)] rescale  the spectrum by a $(1/2)$-shrinking  
\item[(ii)] duplicate the resulting spectrum
\item[(iii)] invert the  duplicate around its center  $\cE = 3\pi/4$. 
\end{itemize}
In Fig.~\ref{fig:self-similarity}(a) this procedure is explicitly illustrated up to level $n=5$. Note that some pairs of adjacent smaller bands merge in  single bands with a  width twice larger. Note also that the inversion  is not trivial only for the first iteration $n=0\to n=1$ since from $n=1$ onward the spectrum becomes symetric around its center.

As a result, for  the $q=2^n$ series, at level $n\ge 1$  there are ${(2^{n}+2)/ 2}$ bands with individual  widths necessarily summing  to 
\beqn
\delta\cE^{\mathrm{tot}}_{2^n} = \frac{2 \pi}{3}.
\label{eq:Cantor:like:2}
\eeqn
By construction, this band spectrum is evidently self-similar, and it becomes  a fractal in the  limit $n \to \infty$. 

Likewise, consider  in Fig.~\ref{fig:self-similarity}(b) the flattened energy bands for the series $q=3^n$  up to level $n=4$. Starting from  level $n$  the iterative procedure consists in 
\begin{itemize}
\item[(i)] rescaling  the  spectrum  by a $(1/3)$-shrinking; 

\item[(ii)] triplicating the resulting spectrum; 
\item[(iii)] and finally inverting the second copy around its center $\cE = \pi/2$  
\end{itemize}
with pairs of adjacent bands again merging into  single bands of width  twice larger. As a result, for  the $q=3^n$ series, at level $n\ge 1$   there are  ${(3^{n}+1)/ 2}$ bands which again have a total width  $\delta\cE^{\mathrm{tot}}_{3^n} =2\pi/3$, as in Eq.~\eq{eq:Cantor:like:2}. Similarly to the $q=2^n$ series, in the limit $n\to\infty$ the $q=3^n$ band spectrum becomes a fractal.

A similar iterative procedure  applies to  the  band spectrum  for the $q=5^n$ series as  illustrated in Fig.~\ref{fig:self-similarity}(c) up to level $n=3$. The bands at  level $n+1$ are constructed from the bands at level $n$ by a  $1/5$-shrinking, $5$-plicating  and inversion of every even copy around their centers  $\cE = 3\pi/10$ and $\cE = 7\pi/10$ respectively.  Again adjacent bands  merge into single   bands twice wider so that at level $n\ge 1$ there are ${(5^n+1)/ 2}$ bands  with a total width $\delta\cE^{\mathrm{tot}}_{5^n} =2\pi/3$. Clearly, in the limit $n\to\infty$ the $q=5^n$ spectrum becomes a fractal.

The universality of the iterative self-similar procedure is now evident. For the $q=m^n$ series  the band spectrum at level $n+1$ can be constructed from the band spectrum at level $n$ by 
\begin{itemize}
\item[(i)] $(1/m)$-shrinking,
\item[(ii)] $m$-plicating,
\item[(iii)] inverting every even copy around their centers  $\cE = (2k-1)\pi/(2m)$ with $k=2,4, \dots$.
\end{itemize}
As a result  at level $n\ge 1$ there are ${(m^n+1)/ 2}$ (for odd $m$) or ${(m^n+2)/ 2}$ (for even $m$) bands with a total width $\delta\cE^{\mathrm{tot}}_{m^n} =2\pi/3$. This implies that the bands always occupy $2/3$ of the available flattened energy range $[0,\pi]$. The universality of the procedure is highlighted by the fact that all the band spectra  are constructed  from the same initial $n=0$ level band spectrum.

For every given base $m$  each successive iteration $n \to n+1$ has the bands shrinking to smaller bands. Thus, in the limit $n \to \infty$, one obtains a point-like   spectrum consisting of  infinitely many bands of infinitely small width. The structure of the band spectrum is similar to a Cantor set with a difference, which is, however,  essential: contrary to the Cantor set, the total band width is the nonzero base-independent quantity
\beqn
\lim_{n\to \infty}\delta\cE^{\mathrm{tot}}_{m^n} = \frac{2 \pi}{3}\,.
\label{eq:self-similaritybis}
\eeqn

Thus, similarly to the original unflattened band spectrum, the flattened band spectrum has a fractal self-similar structure when plotted against the normalised magnetic flux $\beta = p/q$ with $p=1,\dots,q$, with a resolution $q=m^n$  for  base $m$ determined by the increasing level number $n$. The self-similarity pattern depends on $m$ but the total band width~\eq{eq:self-similaritybis}  is $m$-independent. The flattened band spectrum has a Hausdorff dimension $D=1$ which follows from the $1/m$-shrinking $m$-plicating iterative procedure.

Finally, we stress that, as far as the   unflattened spectrum $E(k_y)$ in eq.~\eq{eq:E} is concerned,   distortion effects tend to disappear in the $n\to\infty$ limit: as  $n$ increases the energy  ``ovals'' distortion effects migrate from the center of the spectrum $E \sim 0$ towards its edges,  $E \to \pm 2$. In other words,  in the limit of an infinitely large $q = m^n$ resolution, regions with  distorted ovals   eventually disappear at the edges of the energy domain. This feature is clearly visible in the  plots of Fig.~\ref{fig:2:3:5} which are worth comparing with their corresponding flattened versions in Fig.~\ref{fig:flattened:2:3:5}. In Ref.~\cite{ref:I},  the unflattened $q \to \infty$ spectrum was shown to have a total band width saturating to $8/3$. Compared to the total unflattened energy range, $E \in [-2,2]$, the bands therefore occupy in this limit $2/3$ of the whole energy interval. Equation~\eq{eq:self-similaritybis}, not surprisingly, leads  to the same conclusion for the flattened band spectrum.  

\section{Complex energy branches} 
\label{sec:complex}

So far we have  discussed the real energy branches  of the non-Hermitian Hamiltonian \eq{eq:H:beta}. However, the full spectrum  contains energy branches with a non zero imaginary part. Since these energies correspond either to some formal instabilities or to a dissipative motion of the particles, they are not particularly relevant for a large system at thermodynamical equilibrium. Still, for the sake of  completeness, we find appropriate to display the entire spectrum with its complex energy modes.

\begin{figure}[!htb]
\begin{center}
\includegraphics[width=3.3in]{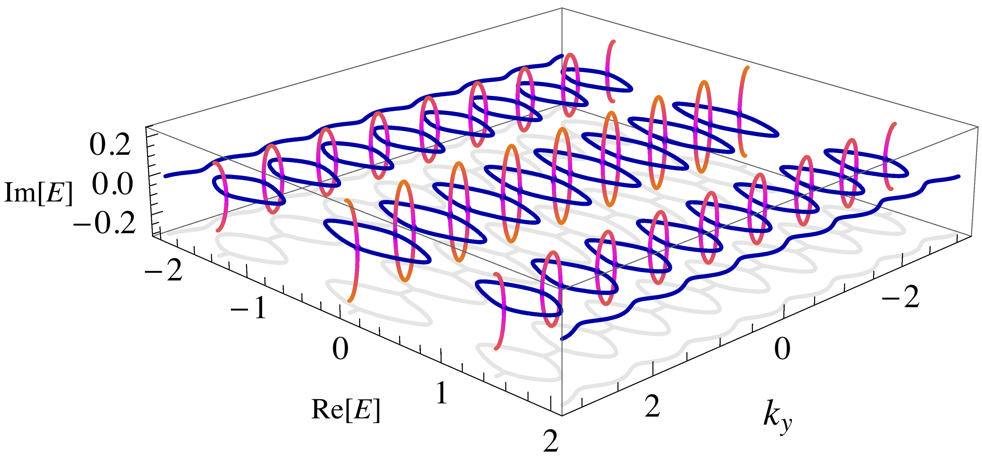}
\end{center}
\vskip -5mm
\caption{The real and imaginary parts of the energy spectrum for $q=8$  plotted against the quasi-momentum $k_y$. The horizontal blue lines represent the real  spectrum ($\IM E = 0$) while the vertical red-orange lines correspond to the complex spectrum with a  nonzero imaginary part ($\IM E \ne 0$). A projection of the energy spectrum on the bottom of the plot is also presented for convenience (grey color).}
\label{fig:3d}
\end{figure}

In Fig.~\ref{fig:3d} the   $q=8$ spectrum is displayed in a 3d ($k_y$, $\RE E$, $\IM E$) plot. The  real branches shown   in Fig.~\ref{fig:real:energy:q8} are recognisable in the horizontal  ($k_y$, $\RE E$, $\IM E =0$) plane. They  build  ordered chains of  distorted ovals. Similarly, the complex branches build  ordered chains of  distorted ovals in the ($k_y$, $\RE E =0$, $\IM E$) plane. These complex (``vertical'' in the 3d plot) ovals connect the real (``horizontal'') ovals in a $3d$ chain--like manner. 

Energy spectra for generic $q$ values, both for periodic and antiperiodic boundary conditions, are similar to  Fig.~\ref{fig:3d}, with horizontal  real ovals always connected by  vertical complex ovals along the $k_y$ coordinate.  This observation gives strong evidence to the fact that the entire energy spectrum -- with the imaginary branches included -- is also a fractal, similarly to the real energy spectrum.

\section{Conclusions}

We have studied the fractal properties of the energy spectrum of a charged particle  with a biased motion on a two-dimensional square lattice  in the background of a magnetic field. This biased quantum model is such that the particle is not allowed to hop in one of the directions (say, to the left) while it can freely move in all  other directions. Its dynamics  is described by a non-Hermitian Hamiltonian of a Hofstadter type.

We have shown that the energy spectrum, which  depends on the normalised magnetic flux -- the ratio of the magnetic flux threading an elementary plaquette to the    magnetic flux quantum -- possesses a nested multi-fractal structure. 

The  energy bands, plotted against the rational magnetic flux, exhibit a fractal pattern shown in Fig.~\ref{fig:butterfly}. Contrary to the Hofstadter's butterfly,  the band spectrum  is rather similar to a spider net.  

The energy levels, plotted against the quasi-momentum, exhibit even more curious fractal patterns as  illustrated in Fig.~\ref{fig:2:3:5}. We have shown, both numerically and analytically, that the  real energy spectrum  is an overlap of infinitely many inequivalent fractals which we call ``fractal energy carpets''. The energy levels in each fractal are space-filling curves with Hausdorff dimensions~$2$. 

In a more rigorous approach, we have shown that for a given base $m$ and resolution $q = m^n$  with $n \in \N$, the energy curves for the normalised magnetic flux $\beta = 1/m^n$    ($m = 2,3,5,6,7,10,\ldots$ is required to not be equal to a natural power of a natural number) have a self-similar geometric structure. This self-similarity is observed up to certain finite scales, both in quasi-momentum and in energy, constrained by the  finiteness of  $n$. In the limit of an infinitely fine resolution, $n \to \infty$, the structure of the energy level becomes a fractal  with various fractal carpets depending on the base  $m$ considered.

We have also shown that the real branches of the energy spectrum are connected by  complex branches  forming chain-like structures in the 3d space of the energies plotted against the quasi-momentum.

The fractal properties of the energy levels are visualised in the supplementary video material~\cite{ref:video}.

\end{document}